\documentclass[12pt]{iopart}

\usepackage{setstack}
\usepackage{graphicx}
\usepackage{amssymb}
\usepackage{bm}
\usepackage{cite}
\usepackage[implicit=false]{hyperref}

\def\be{\begin{equation}}
\def\ee{\end{equation}}
\def\bea{\begin{eqnarray}}
\def\eea{\end{eqnarray}}
\newcommand{\matr}[4]{{\left(\begin{array}{cc} #1&#2\\#3&#4\\\end{array}\right)}}
\newcommand{\vect}[2]{{\left(\begin{array}{c} #1\\#2\\\end{array}\right)}}
\renewcommand{\vec}{\mathbf}

\newcommand{\vr}{\vec{r}}

\newcommand{\vd}{\vec{d}}
\newcommand{\vq}{\vec{q}}

\newcommand{\vX}{\vec{X}}
\newcommand{\ve}{\vec{e}}
\newcommand{\vsigma}{\mbox{\boldmath $\sigma$}}

\newcommand{\vnabla}{\mbox{\boldmath $\nabla$}}

\newcommand{\vp}{\vec{p}}

\renewcommand{\Im}{\mathrm{Im}}

\newcommand{\vF}{v_{\rm F}}
\newcommand{\pF}{p_{\rm F}}

\begin{document}
\title[Quantum corrections to transport in graphene: a semiclassical analysis]{Quantum corrections to transport in graphene: a trajectory-based semiclassical analysis}

\author{Martin Schneider and Piet W. Brouwer}
\address{
Dahlem Center for Complex Quantum Systems and Institut f\"ur Theoretische Physik,
Freie Universit\"at Berlin, Arnimallee 14, 14195 Berlin, Germany
}
\date{\today}

\begin{abstract}
We review a calculation of the quantum corrections to electrical transport in graphene using the trajectory-based semiclassical method. Compared to conventional metals, for graphene the semiclassical propagator contains an additional pseudospin structure, which influences the results for weak localization, and interaction-induced effects, such as the Altshuler-Aronov correction and dephasing. Our results apply to a sample of graphene that is doped away from the Dirac point and subject to a smooth disorder potential, such that electrons follow classical trajectories. In such system, the Ehrenfest time enters as an additional timescale.
\end{abstract}

\maketitle

\section{Introduction}

The discovery of a method to isolate single layers of graphene \cite{novoselov2004} has created a field of intense research over past years (see Refs.\ \cite{castroneto2009,geim2009,beenakker2008,peres2010,dassarma2011} for reviews). The remarkable electronic properties of graphene stem from a quasirelativistic dispersion with a fourfold degeneracy due to spin and valley degrees of freedom, that is characteristic for the underlying two-dimensional honeycomb lattice \cite{wallace1947}. For pristine graphene, the Fermi level lies at the Dirac point, where the density of states vanishes. Electric transport is then dominated by evanescent modes, giving rise to a finite minimum conductivity $4e^2/\pi h$ \cite{fradkin1986,ludwig1994,ziegler1998,shon1998,tworzydlo2006,peres2006,katsnelson2006b}. Remarkably, when smooth disorder is added, that does not couple the valleys, no Anderson localization takes place \cite{bardarson2007,nomura2007,sanjose2007,schuessler2009}, but instead the conductivity rises 
with disorder strength \cite{bardarson2007,nomura2007,dassarma2012}.

When graphene is doped away from the Dirac point, its physical properties resemble those of a metal, but certain intriguing features from the Dirac spectrum remain. The reason for this lies in the ``pseudospin'' degree of freedom, which is connected to the two-atom basis of the two-dimensional honeycomb lattice. Sufficiently close to the Dirac point, where the electronic dispersion is still linear, the direction of the pseudospin is aligned with the momentum. This helicity of charge carriers strongly influences the electronic properties. Two important consequences are the absence of backscattering at a potential barrier (Klein tunneling) \cite{cheianov2006,katsnelson2006} and the half-integral quantum Hall effect \cite{novoselov2005,zhang2005}.

The type of disorder \cite{ostrovsky2006,aleiner2006} and the dielectric environment \cite{nomura2007b,nomura2006,adam2007} play an important role for the transport properties of graphene. When charged impurities in the substrate are the main source of disorder, the disorder potential is smooth on the scale of the lattice constant. For graphene doped away from the Dirac point, one may eventually reach a situation, where the spatial scale $\xi$ on which the disorder changes is much larger than the Fermi wavelength $\lambda_{\rm F}$. (Strictly speaking, this condition is met only if the graphene sheet is embedded in an insulating medium with a high dielectric constant, such as HfO$_2$ \cite{jang2008,ponomarenko2009,newaz2012}.) In this limit, it is justifiable to work with the semiclassical approximation, and to utilize classical trajectories for the calculation of physical quantities. In the past years, trajectory-based semiclassical methods have been successfully applied to the calculation of quantum 
corrections to 
the 
transport of normal metallic systems (see, {\em e.g.}, \cite{aleiner1996,altland2007,schneider2013,richter2002,mueller2009b,brouwer2007b}). The present article discusses the application of such methods to quantum transport in graphene. 

In the limit that the Fermi wavelength $\lambda_{\rm F}$ is much smaller than the transport mean free path $l_{\rm tr}$, the main contribution to the conductivity is given by the Drude conductivity, for which the quantum phase coherence of the electrons plays no role \cite{adam2007,hwang2007}. Corrections to the Drude conductivity are called ``quantum corrections''. They are the result of quantum interference of electrons moving along different trajectories. The two quantum corrections to the conductivity are the weak localization correction, which results from interference of time-reversed trajectories \cite{anderson1979,gorkov1979,hikami1980,lee1985a,chakravarty1986}, and the Altshuler-Aronov correction, which is interaction-induced and originates from elastic scattering of electrons on Friedel oscillations of the electron density \cite{altshuler1980b,altshuler1979c,zala2001}. 
Inelastic interaction processes cause a loss of phase coherence, and set an upper limit on the timescale at which weak localization occurs \cite{altshuler1982b,aleiner1999,marquardt2007}. Although the quantum corrections are a large factor $\sim l_{\rm tr}/\lambda_{\rm F}$ smaller than the Drude conductivity, they can be experimentally detected by their characteristic dependence on magnetic field and temperature. The measurement of such quantities provides important information about the type of disorder and interactions for a specific sample.

Soon after the initial experiments, the theory of quantum corrections in disordered metals was extended to graphene \cite{khveshchenko2006,mccann2006,morpurgo2006,aleiner2006,altland2006,kozikov2010,jobst2012}. Of particular interest was the potential valley-mixing effect of short range disorder (correlation length $\xi \ll \lambda_{\rm F}$), which leads to a transition between weak localization and weak antilocalization \cite{khveshchenko2006,mccann2006,morpurgo2006,aleiner2006} and strongly affects the magnitude of the Altshuler-Aronov correction \cite{kozikov2010,jouault2011,jobst2012}. Such transitions were also observed experimentally \cite{morozov2006,tikhonenko2008,ki2008,tikhonenko2009,cao2010,kozikov2010,chen2010b,jouault2011,laraavila2011,jobst2012}. In the present article, we consider the quantum corrections to the conductivity of graphene in the presence of a {\em long-range} impurity potential, which is smooth on the scale of the Fermi wavelength $\lambda_{\rm F}$. Such a smooth random potential 
does not mix the valleys, but instead leads to a 
number of modifications of the quantum corrections because of its smoothness. Such modifications are known from the theory of conventional electron gases \cite{aleiner1996,yevtushenko2000,brouwer2007b,altland2007,schneider2013}. It is the goal of the present article to extend and collect those results for the case of graphene.

In order to understand why a smooth random potential modifies quantum corrections, one first recalls that electrons follow well-defined classical trajectories if the potential is smooth. The randomness of the potential then ensures that 
the classical dynamics is chaotic: Two nearby trajectories separate exponentially with time, the divergence rate being described by a Lyapunov coefficient $\lambda$. For quantum interference corrections this exponential divergence then leads to the
notion of the ``Ehrenfest time'' $\tau_{\rm E}$ \cite{aleiner1996}:
The Ehrenfest time is defined as the time after which two classical trajectories, initially a quantum distance $\lambda_{\rm F}$ apart, get separated by a reference distance $L_{\rm c}$ characteristic of the classical dynamics
\begin{equation}
  \tau_{\rm E}= \frac{1}{\lambda}
  \ln(L_{\rm c}/\lambda_{\rm F}).
\end{equation}
The significance of the Ehrenfest time is that it puts a short-time cutoff for the occurrence of quantum corrections, as it sets the minimum time for interference effects to take place. In the case of a non-smooth potential with variations on the scale of the Fermi wavelength, no such short-time cutoff appears, and the results of the semiclassical theory agree with those of standard diagrammatic methods.

The trajectory-based semiclassical calculation of quantum corrections to the conductivity of graphene differs from the same calculation for conventional metals by the additional pseudospin structure. The problem of extending the semiclassical formalism to system with a spinor degree of freedom, such as metals with spin-orbit coupling or Dirac Hamiltonians, has received considerable attention in the literature \cite{rubinow1963,littlejohn1991,littlejohn1992,bolte1998,bolte1999,zaitsev2005,zaitsev2005b,silvestrov2006}. The application of the formalism to the case of graphene by Carmier and Ullmo \cite{carmier2008} will serve as a starting point for our calculation. A key element in the trajectory-based approaches is that the pseudospin can be reconstructed along the trajectories, where it remains aligned with the momentum. Associated with the transport of the pseudospin along the trajectory is an additional phase in the semiclassical propagator, which can be identified as the Berry phase. One example where 
this 
phase plays an 
important 
role is the semiclassical calculation of the Landau levels, where the electrons acquire a Berry phase of $\pi$ during the cyclotron motion, ultimately leading to the half-integral quantum Hall effect.

The semiclassical theory presented here is specifically aimed at the leading order quantum corrections (weak localization and Altshuler-Aronov correction, as well as the effect of dephasing on weak localization) for graphene in a smooth random potential. Typical systems to which the trajectory-based semiclassical method has been applied in the literature are quantum billiards, ultraballistic systems, where particles scatter only at the boundary of the sample, or antidot arrays, high mobility two-dimensional electron gases with artificially superimposed antidots, that act as classical scatterers \cite{roukes1989,ensslin1990,yevtushenko2000}.
While for standard semiconductor structures, quantum billiards can be shaped by means of gate potentials, such procedure is problematic for graphene, as it is a gapless material. Quantum billiards in graphene can be realized in etched structures, where the edges are atomically sharp  \cite{ponomarenko2008,schnez2009,moriyama2009,guettinger2009}. The scattering on such edges then depends on the precise atomic configuration, and deserves a careful consideration \cite{wurm2011a, wurm2011b}. The same applies to antidot arrays in graphene \cite{eroms2009}, but also here, the boundaries of the antidots are so sharp, that they lead to scattering between the valleys. Such atomically sharp boundaries invalidate a description that is solely based on classical trajectories, although for sufficiently well-defined boundaries a theoretical description involving coupled valleys is possible \cite{wurm2011a,wurm2011b}. Such a limitation does not exist for the effect of an impurity potential in a graphene sheet on a substrate 
with 
high dielectric constant, which is the scenario we consider here. In this case, the high dielectric constant ensures that the screening length is larger than $\lambda_{\rm F}$ at sufficiently high doping. Hence, the potential has a correlation length $\xi\gg \lambda_{\rm F}$ and classical paths are well-defined objects. 

For the sake of readability, we have tried to make this article self contained, only referring to the literature for the finer technical details of the semiclassical approach. Starting point of our discussion is the semiclassical Green function, that will be introduced in Sec.\ \ref{sec:SemiGreen}. To set the stage, we then calculate the Drude conductance in Sec.\ \ref{sec:Drude}. We then turn to the quantum corrections, where the weak localization is calculated in Sec.\ \ref{sec:WAL}, and the interaction-induced corrections are treated in Sec.\ \ref{sec:AA} (Altshuler-Aronov) and \ref{sec:Dephasing} (dephasing). We conclude in Sec.\ \ref{sec:conclusion}.

\section{Semiclassical Green function}
\label{sec:SemiGreen}

In this article, we consider graphene subject to a smooth disorder potential $V(\vr)$ that does not couple the valleys. In the vicinity of the Dirac point, electrons are described by the Hamiltonian
\begin{equation}
 \label{eq:Hamiltonian}
 H=v_{\rm F} \vp \cdot \vsigma + V(\vr) - \mu, 
\end{equation}
where $v_{\rm F}$ is the Fermi velocity, $\mu$ is the chemical potential, and $\vsigma=(\sigma_x,\sigma_y)$ are Pauli matrices for the pseudospin degree of freedom. The eigenvalues of the kinetic energy term $v_{\rm F} \vp \cdot \vsigma$ are $K_{\pm} = \pm v_{\rm F} |\vp|$. We will be interested in the case of electron-doped graphene for which the chemical potential $\mu$ is larger than the potential $V(\vr)$. In this case we may restrict our attention to the conduction band and set $K = v_{\rm F} |\vp|$. The corresponding eigenspinor of the kinetic energy is
\begin{equation}
 \label{eq:spinor}
 |\chi(\vp)\rangle=\frac{1}{\sqrt{2}}\vect{1}{e^{i \phi_{\vp}}},
\end{equation}
where the angle $\phi_{\vp}$ denotes the direction of the momentum $\vp$.

Starting point for our semiclassical analysis of transport is the semiclassical expression for the retarded Green function ${\cal G}^{\rm R}(\vr,\vr';\varepsilon)$ at energy $\varepsilon$ derived by Carmier and Ullmo \cite{carmier2008},
\begin{equation}
 \label{eq:SemiGreen}
  {\cal G}^{\rm R}(\vr,\vr';\varepsilon)=\frac{2\pi}{(2\pi i\hbar)^{3/2}}\sum_{\alpha:\vr'\rightarrow\vr;\varepsilon} A_{\alpha} e^{i \mathcal{S}_{\alpha}/\hbar + i \gamma_{\alpha}} |\chi(\vp_{\alpha})\rangle\langle\chi(\vp'_{\alpha})|,
\end{equation}
where the summation is over all classical trajectories $\alpha$ that connect the points $\vr$ and $\vr'$, ${\cal S}_{\alpha}$ is the classical action of the trajectory, $A_{\alpha}$ the stability amplitude, $\gamma_{\alpha}$ an additional phase shift to be defined below, and $\vp_{\alpha}'$ and $\vp_{\alpha}$ are the initial and final momenta of the trajectory $\alpha$, respectively. Equation (\ref{eq:SemiGreen}) generalizes the corresponding expression for a system without spin or pseudospin degrees of freedom \cite{gutzwiller1990}. (In that case the projection factor $|\chi(\vp_{\alpha})\rangle\langle\chi(\vp'_{\alpha})|$ and the phase shift $\gamma_{\alpha}$ are absent.) The classical trajectories are determined by the classical Hamilton function
\begin{equation}
 \label{eq:Hclass}
 H_{\rm cl}(\vp,\vr)=v_{\rm F} |\vp|+V(\vr)-\mu,
\end{equation}
and the classical action of a trajectory satisfies the equations
\begin{equation}
 \label{eq:partialS}
  \frac{\partial \mathcal{S}_{\alpha}}{\partial \vr}=\vp_{\alpha},\ \
  \frac{\partial \mathcal{S}_{\alpha}}{\partial \vr'}=-\vp'_{\alpha},  \ \ 
  \frac{\partial \mathcal{S}_{\alpha}}{\partial \varepsilon}=\tau_{\alpha},
\end{equation}
where 
$\tau_{\alpha}$ is the duration of the trajectory. 
The stability amplitude $A_{\alpha}$ is found as $\sqrt{|\det D_{\alpha}|}$, with
\begin{equation}
  D_{\alpha} = \left( \begin{array}{cc} \frac{\partial^2 {\cal S}_{\alpha}}{\partial \vr' \partial \vr} &  \frac{\partial^2 {\cal S}_{\alpha}}{\partial \vr' \partial \varepsilon} \\ \frac{\partial^2 {\cal S}_{\alpha}}{\partial \varepsilon \partial \vr} &  \frac{\partial^2 {\cal S}_{\alpha}}{\partial \varepsilon^2} \end{array} \right).
\end{equation}
Finally, the phase shift $\gamma_{\alpha}$ contains the Berry phase
\begin{equation}
 \gamma_{\alpha}
 \label{eq:Berry}
  =- \frac{1}{2} \int_{0}^{\tau_{\alpha}} dt \frac{d\phi_{\vp_{\alpha}(t)}}{dt} =
  -\frac{1}{2}\left(\phi_{\vp_{\alpha}}-\phi_{\vp'_{\alpha}}+2\pi n\right),
\end{equation}
where in the second equality we chose the initial and final angles to be $0\leq \phi_{\vp'_{\alpha}},\phi_{\vp_{\alpha}}\leq 2\pi$ and add $2\pi n$, $n$ being integer, to account for the phase winding along the path $\alpha$. We have not included the phase corresponding to the Maslov index \cite{gutzwiller1990}, which can be disregarded for the calculation of transport.

In the next Sections we also need the advanced Green function, which follows from the relation
\begin{equation}
 \label{eq:retadv}
   {\cal G}^{\rm A}(\vr',\vr;\varepsilon)=\left[{\cal G}^{\rm R}(\vr,\vr';\varepsilon)\right]^{\dagger}.
\end{equation}

\section{Drude conductance}
\label{sec:Drude}

We now turn to the calculation of the conductivity. Hereto, we consider a rectangular sample of graphene of dimensions $L\times W$, calculate its conductance $G$, and obtain the conductivity $\sigma$ from the relation $G = \sigma W/L$. The conductance $G$ is calculated from the Kubo formula
\begin{eqnarray}
 \label{eq:Kubo}
  G &=&\frac{e^2 d_{\rm g} \hbar}{2\pi} \int_0^W dy  dy' \int d\varepsilon \left(-\frac{\partial f(\varepsilon)}{\partial\varepsilon}\right)
\nonumber\\    && \mbox{}\times
\mathrm{Tr}\left[\hat{v}_{x}{\cal G}^{\rm R}(\vr,\vr';\varepsilon) \hat{v}'_{x} {\cal G}^{\rm A}(\vr',\vr;\varepsilon)\right]_{x'=0,\, x=L},
\end{eqnarray}
where $f(\varepsilon)=1/(e^{\varepsilon/T}+1)$ is the Fermi function and $d_{\rm g}=4$ denotes the degeneracy due to spin and valley. Further, the velocity operator for graphene reads
\begin{equation}
 \hat{v}_x=v_{\rm F} \sigma_x
\end{equation}
and the trace indicates a summation over pseudospin indices.

For a semiclassical calculation of the conductance, we insert the semiclassical Green function \eref{eq:SemiGreen} into the Kubo formula, so that $G$ is expressed as a double sum over trajectories $\alpha$ and $\beta$. Restricting the summation to diagonal terms $\alpha=\beta$, the so-called diagonal approximation, then gives the Drude conductance. For $\alpha=\beta$ the semiclassical approximation Eq.\ (\ref{eq:Kubo}) contains matrix elements of the form
\begin{equation}
 \label{eq:velocity}
 \langle \chi(\vp_{\alpha})|\hat{v}_x|\chi(\vp_{\alpha})\rangle=v_{\rm F} \cos \phi_{\vp_{\alpha}} = v_x,
\end{equation}
with $v_x={\partial H_{\rm cl}}/{\partial p_x}$. Apart from the factor $d_{\rm g}$, the resulting expression is the same as in the case of a standard two-dimensional electron gas,
\begin{equation}
  G_0=d_{\rm g}\frac{e^2}{(2\pi\hbar)^2} \int dy \int dy' \int d\varepsilon \left(-\frac{\partial f(\varepsilon)}{\partial\varepsilon}\right)\sum_{\alpha:(0,y')\rightarrow(L,y)} A_{\alpha}^2  v'_{x} v_x,
\end{equation}
with the initial (final) classical velocity $v'_{x}$ ($v_x$).

The remaining summation over trajectories $\alpha$ can be transformed to an integral over initial and final momentum and the duration of the trajectories \cite{argaman1995,argaman1996},
\begin{eqnarray}
 \label{eq:sumrule}
  \sum_{\alpha:\vr'\rightarrow \vr;\varepsilon} A_{\alpha}^2 f(\vp'_{\alpha},\vp_{\alpha},\tau_{\alpha})
 =\int_{0}^{\infty} dt \int d\vp'_{\varepsilon} \int d\vp_{\varepsilon} 
  \rho_{\varepsilon}(\vX'\rightarrow\vX;t) f(\vp',\vp,t). ~~~
\end{eqnarray}
Here, $f$ is an arbitrary function, and the integration over momenta is restricted to the energy shell $d\vp_{\varepsilon}=d\vp\delta(\varepsilon-H_{\rm cl}(\vr,\vp))$.
The trajectory density $\rho_{\varepsilon}(\vX'\rightarrow\vX;t)$ selects only those phase space points $\vX = (\vr,\vp)$ that are connected via a classical trajectory with the initial point $\vX'$. Since the initial phase space point together with the classical Hamilton function uniquely determines the classical trajectory, $\rho$ is expressed as a $\delta$-function
\begin{eqnarray}
  \rho(\vX'\rightarrow\vX;t) &=&
  \rho_{\varepsilon}(\vX' \rightarrow \vX;t)  \delta[H_{\rm cl}(\vX)-H_{\rm cl}(\vX')] =
  \delta[\vX-\vX(\vX',t)],
\end{eqnarray}
where $\vX(\vX',t) = (\vr(\vr',\vp';t),\vp(\vr',\vp';t))$ is the phase space point that a trajectory starting out of $\vX'=(\vr',\vp')$ reaches after a time $t$.

The calculation proceeds by performing a statistical average of the conductance, where we average over small displacements in the disorder potential $V(\vr)$. Such procedure replaces the exact trajectory density $\rho_{\varepsilon}(\vX'\rightarrow\vX;t)$ by the classical propagator $P(\vX,\vX';t)$, which is a smooth function of initial and final phase space coordinates as well as time. The classical propagator has only a weak dependence on energy which we neglect in the following. We then obtain for the Drude conductance
\begin{eqnarray}
 \label{eq:DrudeSemiclassic}
 G_0 &=&  \frac{e^2 d_{\rm g}}{(2\pi\hbar)^2}\int_0^{\infty} dt \int dy dy'
\int d\vp_{\varepsilon} d\vp'_{\varepsilon}\left[v_{x} P(\vX,\vX';t) v'_{x}\right]_{x'=0,\, x=L}
\end{eqnarray}

For the further analysis of the Drude conductance, one needs to specify the classical propagator $P(\vX,\vX';t)$. Following \cite{bardarson2007,adam2009}, we consider a random Gaussian potential with the correlation function
\begin{equation}
\label{eq:Vcorr}
 \langle V(\vr) V(\vr') \rangle=K_0 \frac{(\hbar \vF)^2}{2\pi\xi^2} e^{-|\vr-\vr'|^2/2\xi^2},
\end{equation}
where $\xi$ is the correlation length, and $K_0$ is the dimensionless strength of the potential.
On spatial scales much longer than the correlation length, the electronic motion becomes diffusive, with a diffusion coefficient that we will calculate in the following. The random potential of Eq.\ (\ref{eq:Vcorr}) has been used to describe the impurity potential in experiments \cite{tan2007,adam2008}.

We start by considering a particle that moves in the $x$-direction at time $t=0$ and consider how the direction of motion changes under the influence of the potential $V(\vr)$. Following the classical dynamics, the angle $\phi(t)$, at which the electron propagates at time $t$ is given by
\begin{equation}
 \phi(t)=-\frac{1}{\hbar k_{\rm F}} \int_0^t dt' \partial_y V (\vr(t')),
\end{equation}
where $k_{\rm F}$ is the Fermi wavenumber. This equation is valid for times $t$ short enough, such that $|\phi(t)|\ll 1$, so that the motion of the electron is mainly along the $x$-direction, $\vr(t)=\vF t \ve_x$. We then find for the mean quadratic deflection
\begin{equation}
 \langle \phi(t)^2\rangle=\frac{1}{(\hbar k_{\rm F} \vF)^2} \int_0^{\vF t} dx dx' \langle \partial_y V(x,0) \partial_y V(x',0)\rangle.
\end{equation}
Using the correlation function Eq.\ \eref{eq:Vcorr}, this gives
\begin{equation}
 \label{eq:deflection}
 \langle \phi(t)^2\rangle =\frac{K_0 \vF t}{\xi^3 k_{\rm F}^2 \sqrt{2\pi}},
\end{equation}
provided $t$ is much longer than the ``correlation time'' $t_{\xi}=\xi/\vF$. Our derivation required the time $t$ to be short enough such that the deflection is small. Such time interval exists, as long as $K_0\ll (k_{\rm F}\xi)^2$, which is a condition that can be met if the disorder is smooth on the scale of the Fermi wavelength $(k_{\rm F} \xi \gg 1)$.

Equation \eref{eq:deflection} describes a linear-in-time increase of the quadratic deflection, a characteristic property of the diffusive motion for the angle $\phi$. Continuing the diffusive process beyond small angles, the validity of this equation can be extended to all times longer than the correlation time $t_{\xi}$. Further, extending the result to arbitrary starting times and arbitrary directions in the beginning of the propagation, we conclude that the angle difference $\phi(t)-\phi(t')$ has a Gaussian distribution with zero mean and with variance
\begin{equation}
 \label{eq:phiphi}
 \langle \left[\phi(t)-\phi(t')\right]^2\rangle=\frac{K_0 \vF}{\xi^3 k_{\rm F}^2 \sqrt{2\pi}}|t-t'|.
\end{equation}

Now we can calculate the electron's mean square displacement. Since
\begin{equation}
 \vr(t)-\vr(0)=\vF \int_0^t dt'
 [\ve_x \cos \phi(t') + \ve_y \sin \phi(t')],
\end{equation}
the mean square displacement is given by
\begin{equation}
 \left\langle |\vr(t)-\vr(0)|^2 \right\rangle=\vF^2\int_0^t dt' dt'' 
  \left\langle \cos[\phi(t')-\phi(t'')] \right\rangle.
\end{equation}
The average can be performed using the Gaussian distribution of $\phi(t')-\phi(t'')$ and one finds, in the long-time limit,
\begin{equation}
 \langle\left[\vr(t)-\vr(0)\right]^2\rangle=4D|t|,
\end{equation}
where the diffusion constant $D$ is given by
\begin{equation}
 \label{eq:D}
 D=\frac{\xi^3 k_{\rm F}^2 \vF \sqrt{2\pi}}{K_0}.
\end{equation}
The diffusion constant of Eq.\ (\ref{eq:D}) corresponds to a transport mean free path 
\begin{equation}
  l_{\rm tr} = \tau_{\rm tr} v_{\rm F} =
  \frac{2 \xi^3 k_{\rm F}^2 \sqrt{2\pi}}{K_0},
  \label{eq:mfp}
\end{equation}
which is parametrically larger than the correlation length $\xi$ in the limit  $k_{\rm F} \xi \gg 1$ of a smooth potential.

We now turn back to the evaluation of Eq.\ \eref{eq:DrudeSemiclassic} in the diffusive limit. In this case, the direction of the momentum does not influence the propagation, which amounts to the replacements
\begin{equation}
 d\vp_{\varepsilon}=(2\pi\hbar)^2 \nu \frac{d\phi}{2\pi},
\end{equation}
and
\begin{equation}
 P(\vX,\vX';t)=\frac{1}{(2\pi \hbar)^2 \nu} P(\vr, \vr';t).
\end{equation}
where $\nu$ is the density of states per spin and valley.
The relevant classical propagator $P(\vr, \vr';t)$ is then a function of position only, and satisfies a diffusion equation
\begin{equation}
 \label{eq:diff}
 (\partial_{t}-D\Delta_{\vr})P(\vr, \vr';t)=\delta(t)\delta(\vr-\vr'),
\end{equation}
where $D$ is the diffusion constant, determined above.

The solution to the diffusion equation Eq.\ \eref{eq:diff} in two dimensions, with perfect leads at $x=0$ and $x=L$, and insulating boundaries at $y=0$ and $y=W$ is given by
\begin{equation}
 \label{eq:diffpropagator}
 P(\vr,\vr';t)=\theta(t)\sum_{\vq}\psi_{\vq}(\vr)\psi_{\vq}(\vr') e^{-D\vq^2 t},
\end{equation}
where we introduced the function
\begin{equation}
 \psi_{\vq} (\vr)=\sqrt{\frac{4}{LW}} \sin(q_x x)
\times \cases{1/\sqrt{2} & \mbox{if $q_y=0$}, \\
                                          \cos(q_y y) & \mbox{if $q_y\neq 0$}.}
\end{equation}
The summation over $\vq$ extends over $q_x={n\pi}/{L}$ with $n=1,2,...$ and $q_y={m \pi}/{W}$ with $m=0,1,...$.

In our calculation of the Drude conductance, as well as in the forthcoming calculation of the quantum corrections, we also encounter expressions, where the diffusion propagator that connects to the leads is multiplied by the velocity $v_x$. Such expressions are handled with the help of the diffusive flux 
\begin{equation}
 j_x(\vr,\vr';t)=-D\partial_x P(\vr,\vr';t),
\end{equation}
at position $\vr$ and time $t$, for a particle starting from $\vr'$ at time $t=0$.
We find
 \begin{equation}
 \eqalign{
 \label{eq:PLR} 
  P_{\rm L}(\vr)&= \int_0^{\infty} dt \int  dy'   d\vp'_{\varepsilon}\left[ P(\vX,\vX';t) v'_{x}\right]_{x'=0}=\frac{L-x}{L},\nonumber\\
  P_{\rm R}(\vr) &= \int_0^{\infty} dt \int dy'    d\vp'_{\varepsilon} \left[v'_{x} P(\vX',\vX;t) \right]_{x'=L}=\frac{x}{L},\nonumber\\
  P_{\rm LR}&= \int_0^{\infty} dt \int dy dy'   \frac{d\vp_{\varepsilon} d\vp'_{\varepsilon}}{(2\pi\hbar)^2\nu}\left[v_{x} P(\vX,\vX';t) v'_{x}\right]_{x'=0, \,x=L}
   =D\frac{W}{L}.
 }
\end{equation}
The quantity $P_{\rm LR}$ appears in the Drude conductance, while the quantities $P_{\rm L}$ and $P_{\rm R}$, which represent the probability that a particle at position $\vr$ has entered via the left lead or will exit via the right lead, respectively, are introduced for later use.

For the Drude conductivity $\sigma_0 = G_0 L/W$ we then obtain the standard expression
\begin{equation}
  \sigma_0=d_{\rm g} e^2\nu D,
\end{equation}
where the factor $d_{\rm g}=4$ accounts for the degeneracy for spin and valley. Importantly, pseudospin does not enter as an additional degeneracy, since it is locked to the momentum. Taking the expression for the diffusion coefficient for the Gaussian random potential, Eq.\ \eref{eq:D}, as well as the density of states at graphene, $\nu=k_{\rm F}/2\pi\hbar\vF$, one obtains
\begin{equation}
 \sigma_0=\frac{4 e^2}{h} \frac{(k_{\rm F} \xi)^3}{K_0}\sqrt{2 \pi}.
\end{equation}
The same result was obtained in a quantum-mechanical calculation using the Boltzmann equation in \cite{adam2009}.

\section{Weak antilocalization}
\label{sec:WAL}
Deviations from the Drude conductance are termed quantum corrections. Without interactions and for conventional metals, the leading correction to the classical conductance results in a small reduction of the conductance, and is called ``weak localization'', since it describes the onset of Anderson localization. In graphene, the Berry phase is responsible for a different sign of this quantum correction, which gives rise to an enhanced conductance (when effects of intervalley scattering and trigonal warping are neglected), and is therefore called weak {\it anti}localization \cite{khveshchenko2006,mccann2006,morpurgo2006,tikhonenko2008,tikhonenko2009}. In the following, we will show how the weak antilocalization correction is derived in the semiclassical formalism, and discuss the effect of a finite Ehrenfest time. Again, we will give explicit results for the case of a smooth random Gaussian-correlated potential.

\begin{figure}[t]
\begin{center}
\includegraphics[width=3.4in]{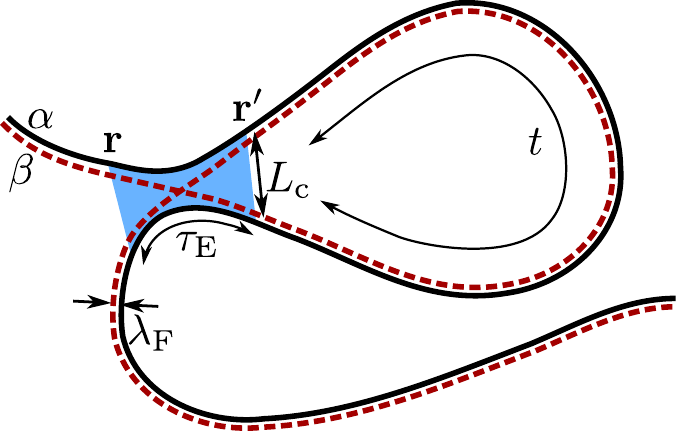}
\end{center}
\caption{Configurations of trajectories $\alpha$ and $\beta$ responsible for weak antilocalization. We represent trajectories corresponding to retarded (advanced) Green functions as solid (dashed) lines. The trajectories consist of entrance/exit segment, a loop segment of variable duration $t$, and an encounter region, which allows for ``pair switching'' of the trajectories (indicated in blue). When trajectories are paired together, their allowed spatial separation is set by the Fermi wavelength $\lambda_{\rm F}$. During the encounter region, this separation gets magnified to a classical size $L_{\rm c}$ beyond which the trajectories develop in an uncorrelated manner. The encounter is shown enlarged in the figure, in reality all four segments of trajectories within the encounter remain very close together, on a submacroscopic scale in phase space.}
\label{fig:WL}
\end{figure}

In the semiclassical framework, weak (anti)localization results from configurations of retarded and advanced trajectories $\alpha$ and $\beta$ as shown in Fig.\ \ref{fig:WL}. The trajectories can be divided into four segments: The entrance and exit segments, where the trajectories $\alpha$ and $\beta$ are correlated or ``paired'' --- {\em i.e.}, the difference between the two trajectories is sufficiently small, that the chaotic classical dynamics can be linearized on that scale ---, the loop segment, where the trajectory $\alpha$ is paired with the time-reversed of trajectory $\beta$, and the encounter region (or Lyapunov region), where trajectories $\alpha$ and $\beta$ as well as their time-reversed are correlated. 
At the beginning of their first passage through the encounter region, the trajectories $\alpha$ and $\beta$ are located within a Fermi wavelength $\lambda_{\rm F}$. Due to the chaotic motion, this phase-space distance increases exponentially along the encounter region as $d(t)=\lambda_{\rm F} e^{\lambda t}$, where $\lambda$ is the Lyapunov coefficient characteristic of the chaotic motion. For the random Gaussian potential (\ref{eq:Vcorr}), the Lyapunov exponent is
\begin{equation}
  \lambda=
  \frac{v_{\rm F}}{\xi \Gamma(1/6)} \left(
  \frac{9 K_0 \pi}{4 (k_{\rm F} \xi)^2 \sqrt{2}} \right)^{1/3},
\end{equation}
see \cite{aleiner1996} and Appendix \ref{sec:applyapunov}.
At the end of the encounter region, the distance has reached a classical size $L_{\rm c}$, beyond which classical motion is considered uncorrelated --- {\em i.e.}, the classical dynamics can no longer be linearized. For the smooth random potential (\ref{eq:Vcorr}), we may identify $L_{\rm c} \simeq \xi$. The duration of the encounter is set by the Ehrenfest time 
\begin{equation}
 \label{eq:tauE}
 \tau_{\rm E}=\lambda^{-1} \ln (L_{\rm c}/\lambda_{\rm F}). 
\end{equation}
Our final results can be expressed in terms of Ehrenfest time only, which depends logarithmically on $L_{\rm c}$, so that a more precise definition of the cutoff $L_{\rm c}$ is not needed.

For the calculation of the weak localization, one starts from the Kubo formula, Eq.\ \eref{eq:Kubo}, inserts the semiclassical expressions for the Green function, and then restricts the summation to configurations of trajectories as explained in the previous paragraph. As long as the duration of the encounter region is $\tau_{\rm E}$ or larger, the trajectories $\alpha$ and $\beta$ acquire an action difference $\Delta \mathcal{S}\lesssim \hbar$ \cite{aleiner1996,richter2002}.

For graphene, we also have to keep track of the influence of the pseudospin, which has two effects: First, the spinor structure of the semiclassical Green function changes the velocity operator to the classical velocity, in the same way as before for the Drude conductance, see Eq.\ \eref{eq:velocity}. Second, since the trajectories are no longer equal, we have to pay attention to the Berry phase collected along the trajectories $\alpha$ and $\beta$. At this stage, we can write
\begin{equation}\eqalign{
  \delta G_{\rm WAL} =&\frac{e^2d_{\rm g}}{(2\pi\hbar)^2} \int dy \int dy' \int d\varepsilon \left(-\frac{\partial f(\varepsilon)}{\partial\varepsilon}\right)\nonumber\\
    & \mbox{}\times\sum_{\alpha,\beta:(0,y')\rightarrow(L,y)} A_{\alpha}^2  v'_{x} v_x e^{i \left(\mathcal{S}_{\alpha}-\mathcal{S}_{\beta}\right)/\hbar} e^{i\left(\gamma_{\alpha}-\gamma_{\beta}\right)},
  \label{eq:dGWAL}
}\end{equation}
where the summation is restricted to the configurations of trajectories shown in Fig.\ \ref{fig:WL}, for which we have $A_{\alpha}=A_{\beta}$.

The difference of the Berry phase $\Delta \gamma=\gamma_{\alpha}-\gamma_{\beta}$, is collected in the loop segment only. (In the encounter region, the trajectories $\alpha$ and $\beta$ differ on a sub-macroscopic scale only, which adds a negligible contribution to the Berry phase difference.) Since the momenta of trajectory $\alpha$ are opposite in the beginning and the end of the loop segment, we have from Eq.\ \eref{eq:Berry}
\begin{equation}
 \gamma_{\alpha,{\rm Loop}}=\pi\left(n+\frac{1}{2}\right),
\end{equation}
with integer $n$ depending on the total winding of the momentum along the trajectory. Since the Berry phase is expressed as integral along the trajectory, the Berry phase collected by trajectory $\beta$ along the loop is just $\gamma_{\beta,{\rm Loop}}=-\gamma_{\alpha,{\rm Loop}}$. Hence, we find
\begin{equation}
 e^{i\Delta \gamma}=-1
\end{equation}
for all configurations of trajectories contributing to the quantum correction. This minus sign is responsible for the change from weak localization (in conventional two-dimensional electron gases without spin-orbit coupling) to weak antilocalization.

\begin{figure}[t]
\begin{center}
\includegraphics[width=3.4in]{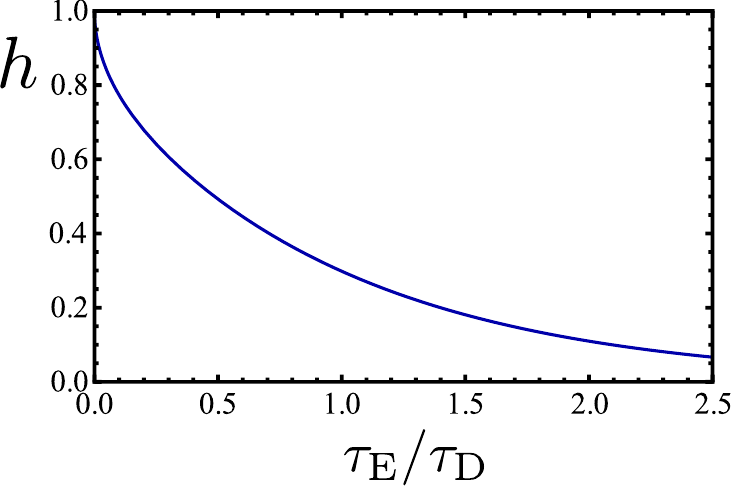}
\end{center}
\caption{At finite Ehrenfest time, weak antilocalization is suppressed by an additional factor $h(\tau_{\rm E}/\tau_{\rm{D}})$}
\label{fig:h}
\end{figure}

The remaining calculation then proceeds as in the standard case, and we find \cite{brouwer2007b}
\begin{equation}\eqalign{
 \label{eq:WALeq}
  \delta G_{\rm WAL}=&-d_{\rm g}\frac{e^2}{2\pi\hbar} \int d\vr d\vr' P_{\rm L}(\vr) P_{\rm R} (\vr) \partial_{\tau_{\rm E}} P(\vr',\vr;\tau_{\rm E}) 
   \int dt P(\vr',\vr';t)
}\end{equation}
where $P(\vr',\vr';t)$ is the diffusion propagator, see Eq.\ \eref{eq:diffpropagator}, and $P_{\rm L}(\vr)$ and $P_{\rm R}(\vr)$ are defined in Eq.\ \eref{eq:PLR}.
For the further evaluation of Eq.\ \eref{eq:WALeq}, we write
\begin{equation}
 \partial_{\tau_{\rm E}} P(\vr',\vr;\tau_{\rm E})=D\Delta_{\vr} P(\vr',\vr;\tau_{\rm E})
\end{equation}
and perform two partial integrations on $\vr$. Making use of the explicit form of $P_{\rm L}(\vr)$ and $P_{\rm R}(\vr)$, we arrive at
\begin{equation}
 \label{eq:WALexpr}
   \delta G_{\rm WAL} = \frac{e^2 d_{\rm g}}{2\pi\hbar} \frac{2D}{L^2}\int d\vr d\vr' P(\vr',\vr;\tau_{\rm E}) \int_{\tau_{\rm tr}}^{\infty} dt P(\vr',\vr';t).
\end{equation}
In two dimensions, the time integral in this equation is divergent for small times, and the appropriate cutoff is set by the transport time $\tau_{\rm tr}$, below which the diffusive approximation breaks down.
In the limit of large aspect ratio $W/L$, and small $\tau_{\rm tr}/\tau_{\rm D}$, where $\tau_{\rm D}=L^2/D\pi^2$ is the dwell time, we then find (see Appendix \ref{sec:appWAL})
\begin{equation}
\label{eq:WALresult}
  \delta \sigma_{\rm WAL}=\frac{e^2d_{\rm g}}{4\pi^2\hbar} \ln(\tau_{\rm D}/\tau_{\rm tr}) h({\tau_{\rm E}}/{\tau_{\rm D}}),
\end{equation}
where the function $h(x)$ is defined as
\begin{equation}
 h(x)=\frac{8}{\pi^2} \sum_{n=1, \, n\,\mathrm{odd}}^{\infty} \frac{1}{n^2} e^{-n^2 x}.
\end{equation}
It has the asymptotic behavior
\begin{equation}
 h(x)=\cases{
       1-\frac{4}{\pi^{3/2}}\sqrt{x}, & $x\ll 1$\\
       \frac{8}{\pi^2} e^{-x},& $x\gg 1$.
      }
\end{equation}
At zero Ehrenfest time, $h(0)=1$ and we arrive at the well-known result for weak antilocalization of a symplectic metal \cite{hikami1980,lee1985a}. At finite Ehrenfest time, our calculation results in a suppression of the weak antilocalization by the additional factor $h(\tau_{\rm E}/\tau_{\rm D})$, shown in Fig.\ \ref{fig:h}. (The same multiplicative factor $h(\tau_{\rm E}/\tau_{\rm D})$ describes the suppression of weak localization or weak antilocalization in a conventional two-dimensional electron gas. We are not aware of a calculation of the function $h$ in this context.)

\section{Altshuler-Aronov correction}
\label{sec:AA}

We now turn to the effects of interactions on the conductivity. Interactions modify the conductivity in two physically distinct ways. First, interactions cause the so-called Altshuler-Aronov correction \cite{altshuler1979c,zala2001}, which has its origin in the interference of elastic scattering off impurities and off Friedel oscillations of the electron density around an impurity. Second, inelastic electron-electron scattering is responsible for a loss of phase coherence or ``dephasing'', which sets an upper limit on the time at which weak (anti)localization can occur.

\subsection{Lowest-order interaction correction}

The semiclassical treatment of interaction corrections proceeds via two steps. First, one considers a specific random potential and includes interactions to first order diagrammatic perturbation theory. Such procedure is rather standard, and results in expressions in terms of Green functions ${\cal G}(\vr,\vr';\varepsilon)$ for the given disorder realization. The second step is to take the disorder average, where we employ the semiclassical framework. Hereto, we insert the semiclassical expressions for the Green functions and identify the relevant configurations of trajectories that contribute to the interaction corrections. Our results take into account the effects of a finite Ehrenfest time.

We now outline the calculation in more detail. For the inclusion of interactions using diagrammatic perturbation theory, we can adapt the results of \cite{aleiner1999}, where the authors divide their expressions into contributions to the Altshuler-Aronov correction and to dephasing effects. For the calculation of the conductance, as done in this work, we may further simplify the expressions of \cite{aleiner1999} by keeping only terms, where one retarded and one advanced Green function is attached to the current vertex (in contrast to the calculation of the conductivity, where also diagrams with two Green functions of the same kind connected to the current vertex need to be kept, see discussion in \cite{schneider2013}).

Explicit expressions for the Altshuler-Aronov conductance correction $\delta G_{\rm AA}$ are obtained from Eq.\ (\ref{eq:Kubo}) upon replacing the retarded Green function ${\cal G}^{\rm R}(\vr,\vr';\varepsilon)$ by ${\cal G}^{\rm R}(\vr,\vr';\varepsilon) + \delta {\cal G}^{\rm R,F}(\vr,\vr';\varepsilon) + \delta {\cal G}^{\rm R,H}(\vr,\vr';\varepsilon)$, and a similar replacement for the advanced Green function ${\cal G}^{\rm A}(\vr,\vr';\varepsilon)$, keeping terms to first order in the interaction only. The functions $\delta {\cal G}^{\rm R,F}(\vr,\vr';\varepsilon)$ and $\delta {\cal G}^{\rm R,H}(\vr,\vr';\varepsilon)$ are Fock and Hartree corrections to the single-particle Green function, respectively,
\begin{eqnarray}
  \label{eq:dGFock}
  \fl\delta {\cal G}^{\rm R,F}_{\alpha\beta} (\vr,\vr';\varepsilon) =\sum_{\gamma\delta}\int \frac{d\omega}{4\pi i}
  \int d\vr_1 d\vr_2\,
  \tanh \left(\frac{\omega-\varepsilon}{2 T}\right)   {\cal G}^{\rm R}_{\alpha\gamma}(\vr,\vr_1;\varepsilon)  {\cal G}^{\rm R}_{\delta\beta}(\vr_2,\vr';\varepsilon)\nonumber\\ \fl
  \qquad \qquad \mbox{} \times \lbrace U^{\rm A}(\vr_1,\vr_2;\omega)  {\cal G}^{\rm R}_{\gamma\delta}(\vr_1,\vr_2;\varepsilon-\omega)
  -U^{\rm R}(\vr_1,\vr_2;\omega)  {\cal G}^{\rm A}_{\gamma\delta}(\vr_1,\vr_2;\varepsilon-\omega)\rbrace,  \\ \nonumber\\
 \label{eq:dGHartree}
  \fl \delta {\cal G}^{\rm R,H}_{\alpha\beta} (\vr,\vr';\varepsilon)=-d_{\rm g}\sum_{\gamma\delta}
  \int \frac{d\omega}{4\pi i} \int d\vr_1 d\vr_2 \tanh \left(\frac{\omega-\varepsilon}{2 T}\right)   {\cal G}^{\rm R}_{\alpha\gamma}(\vr,\vr_1;\varepsilon)  {\cal G}^{\rm R}_{\gamma\beta}(\vr_1,\vr';\varepsilon)\nonumber\\ \fl
  \qquad \qquad \mbox{} \times \lbrace U^{\rm A}(\vr_1,\vr_2;0)  {\cal G}^{\rm R}_{\delta\delta}(\vr_2,\vr_2;\varepsilon-\omega) - U^{\rm R}(\vr_1,\vr_2;0)  {\cal G}^{\rm A}_{\delta\delta}(\vr_2,\vr_2;\varepsilon-\omega)\rbrace.
\end{eqnarray}
In these expressions we wrote the pseudospin indices explicitly. We further allow for a frequency dependence of the interaction propagator $U^{\rm R}(\vr_1,\vr_2;\omega)$ to include the effect of dynamical screening. 
To first order in interaction we also obtain additional corrections to the conductance which are contributing to dephasing. These will be discussed in the next section.

\begin{figure}[t]
\begin{center}
\includegraphics[width=3.5in]{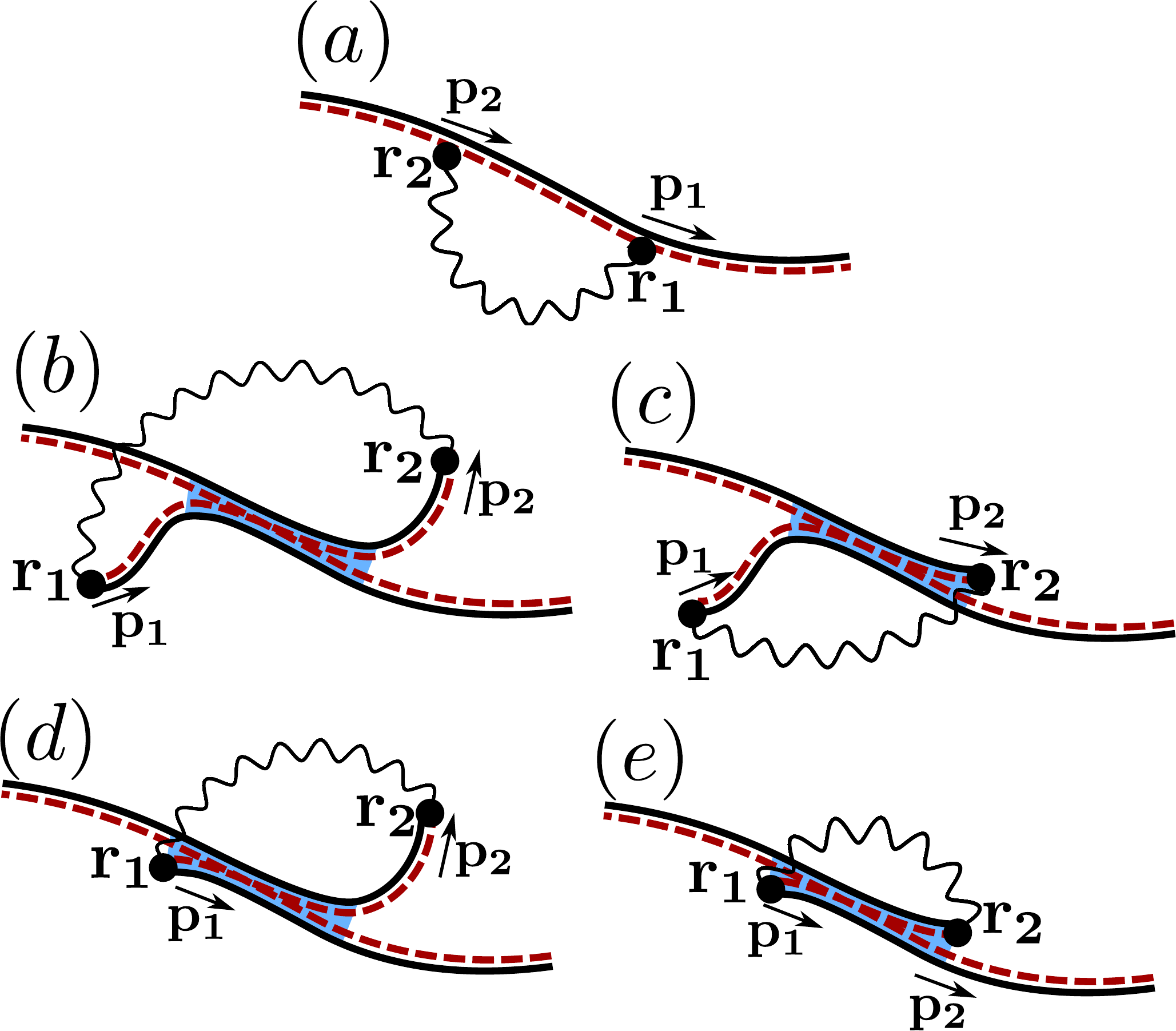}
\end{center}
\caption{Configurations of trajectories that contribute to the Fock contribution to the Altshuler-Aronov correction. ``Retarded'' and ``advanced'' trajectories are represented by solid and dashed lines, respectively. Encounter regions are indicated in blue.}
\label{fig:Fock}
\end{figure}

Insertion of the semiclassical Green functions leads to a sum over four trajectories. Systematic contributions to $\delta G_{\rm AA}$ are obtained only if trajectories originating from retarded and advanced Green functions are paired up or if they undergo a small-angle encounter \cite{schneider2013}. Configurations of trajectories that are in line with these requirements for the Fock contribution to $\delta G_{\rm AA}$ are shown in Fig. \ref{fig:Fock}. Here, configuration $(a)$ originates from a term with three advanced Green functions and one retarded Green function. In this case, the three ``advanced'' trajectories must join to a single trajectory that can be paired up with the ``retarded'' trajectory. Configurations $(b)$-$(e)$ correspond to a term with two retarded and two advanced trajectories. In this situation, due to the specific requirements on start and end point of the Green functions, the trajectories cannot be paired one by one, instead the four trajectories undergo a small-angle encounter. The 
subdivision into configurations $(b)$-$(e)$ reflects the possibilities to have none, one or both interaction points within the encounter region. For each one of the configurations shown in Fig. \ref{fig:Fock}, there is a counterpart for which the role of retarded and advanced trajectories is interchanged. 

In close analogy to the calculation for conventional electron gases \cite{schneider2013}, we find, for a random potential with Gaussian correlations as in Eq.\ (\ref{eq:Vcorr}),
\begin{equation}\eqalign{
 \label{eq:dGAA-F}
 \delta G_{\rm AA}^{\rm F}=&\frac{d_{\rm g} \nu e^2}{2 \pi \hbar^2}  \int d\omega \frac{\partial}{\partial \omega} \left(\omega \coth\frac{\omega}{2 T}\right) \int d\vr_1 d\vr_2 \nonumber\\
   &\times \mathrm{Im}\left\lbrace   U^{\rm R}(\vr_1,\vr_2;\omega) K (\vr_1,\vr_2;\omega) 
   \langle\Sigma_{\rm F}(\vp_1,\vp_2)\rangle_{\vp_1,\vp_2} \right\rbrace,
}\end{equation}
with the kernel $K(\vr_1,\vr_2;\omega)$ given by

\begin{equation}\eqalign{
  \label{eq:KtauE}
  K(\vr_1,\vr_2;\omega)=&-\frac{4D^2}{L^2} \int d\vr d\vr' P(\vr,\vr_1;\omega) P(\vr_2,\vr';\omega)\nonumber\\
                    &\times\partial_x \partial_{x'} \int_{\tau_{\rm E}}^{\infty} dt P(\vr',\vr;t) e^{i\omega t/\hbar},
}\end{equation}
where 
\begin{equation}
 P(\vr,\vr';\omega)=\int_0^{\infty} dt P(\vr,\vr;t) e^{i\omega t/\hbar}
\end{equation}
is the Fourier transform of the diffusion propagator $P(\vr,\vr';t)$, and where the brackets $\langle \ldots \rangle_{\vp_1,\vp_2}$ denote an average over the directions of the momenta $\vp_1$ and $\vp_2$. Equation (\ref{eq:KtauE}) demonstrates that the Ehrenfest time is acting as a short-time cutoff for the Altshuler-Aronov correction.

\begin{figure}[t]
\begin{center}
\includegraphics[width=3.8in]{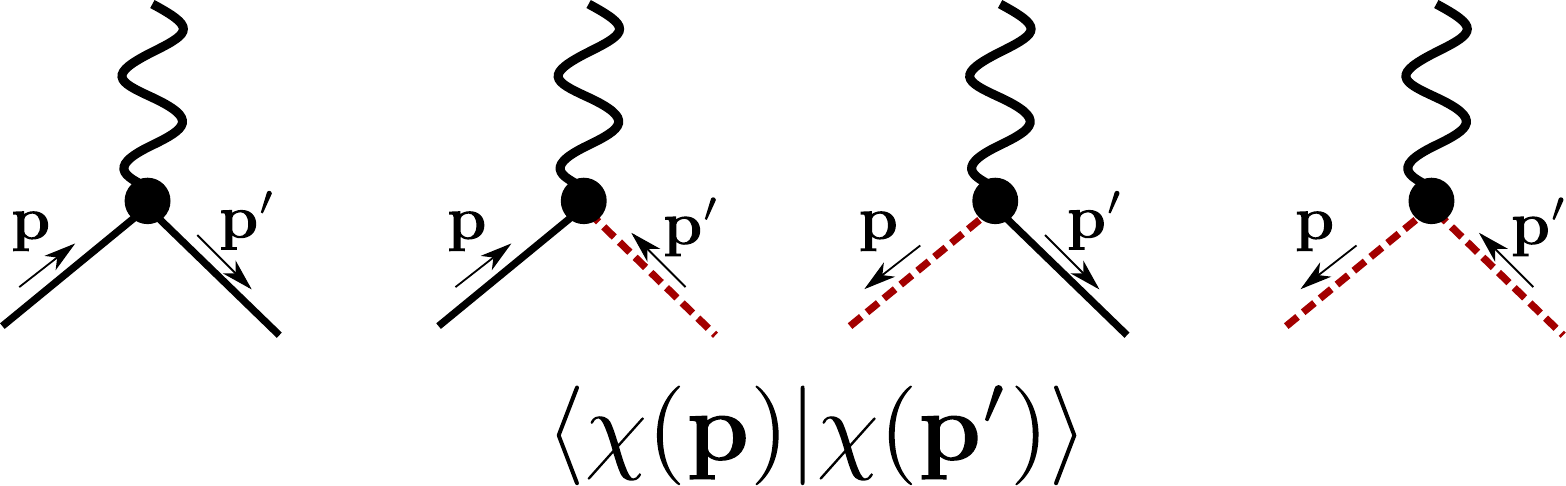}
\end{center}
\caption{For graphene, the pseudospin structure contributes additional factors associated with the interaction vertex. In our notation, the retarded Green function $\mathcal{G}^{\rm R}(\vr_2,\vr_1)$ is associated with trajectories running from $\vr_1$ to $\vr_2$, while the advanced Green function $\mathcal{G}^{\rm A}(\vr_2,\vr_1)$ is associated with trajectories running from $\vr_2$ to $\vr_1$. This amounts to the following possibilities: If the interaction vertex is associated with two Green functions of the same kind, then one trajectory is pointing towards the vertex, while the other one is pointing away. For an interaction vertex associated with one retarded and one advanced Green function, both trajectories either point towards or away from the vertex. In the figure, we show four possibilities that result in a factor $\langle\chi(\vp)|\chi(\vp')\rangle$, with the associated labelling of momenta.}
\label{fig:vertex}
\end{figure}

The spinor structure of the semiclassical Green function contributes the factor $\Sigma_{\rm F} (\vp_1,\vp_2)$, which is not present in the calculation for the conventional two-dimensional electron gas. As explained in Fig.\ \ref{fig:vertex}, it depends on the overlap of pseudospinors of the two trajectories at the interaction points. For the diffusive motion, only the quantity averaged over momenta $\vp_1$ and $\vp_2$ with $|\vp_1|=|\vp_2|=p_{\rm F}$ is relevant. For the Fock contribution, momentum does not change at the interaction points, and the spinors at the interaction points combine to a factor
\begin{equation}
 \Sigma_{\rm F}(\vp_1,\vp_2)=\langle\chi(\vp_1)|\chi(\vp_1)\rangle \langle\chi(\vp_2)|\chi(\vp_2)\rangle=1.
\end{equation}
Thus, up to the degeneracy $d_{\rm g}$, the result for the Fock contribution remains unchanged as compared to the conventional metal. The spinor structure will however influence the Hartree contribution $\delta G_{\rm AA}^{\rm H}$, as we now show.

\begin{figure}[t]
\begin{center}
\includegraphics[width=3.5in]{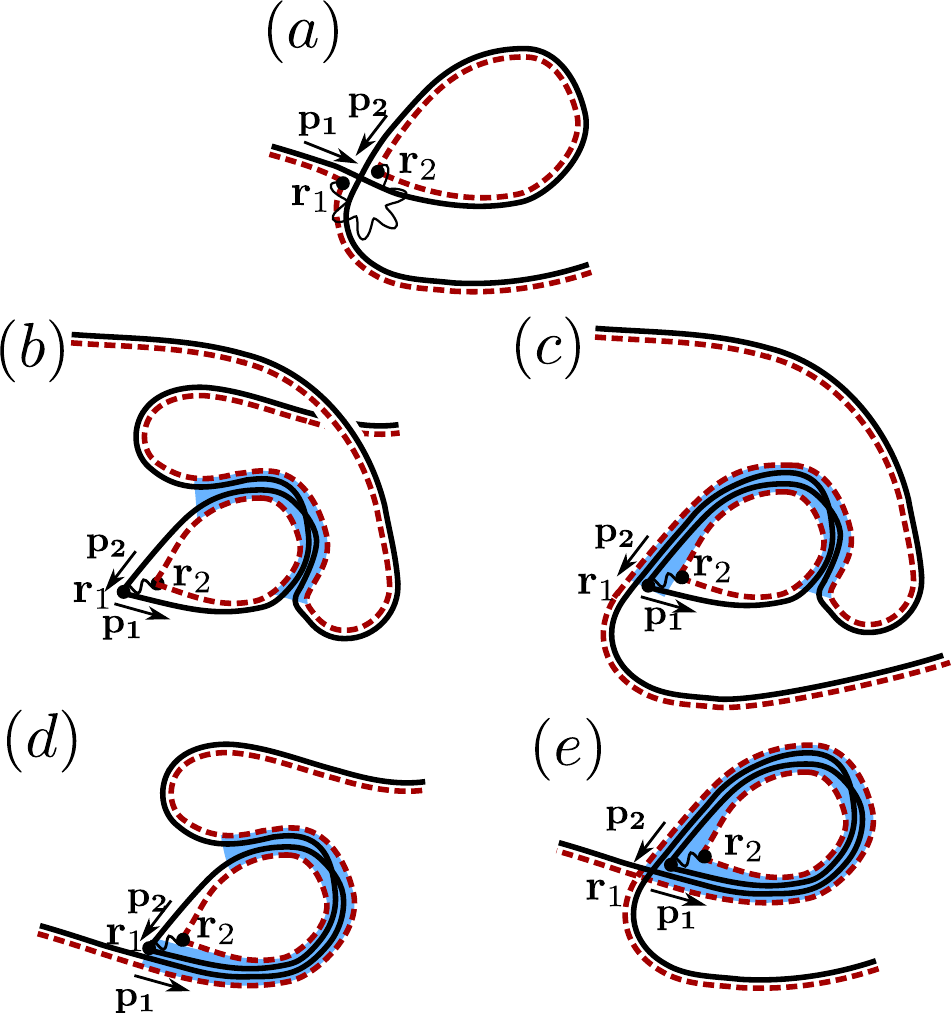}
\end{center}
\caption{Configurations of trajectories that contribute to the Hartree contribution to the Altshuler-Aronov correction. There is a one-to-one correspondence between the configurations for the Hartree and the Fock contribution, Fig. \ref{fig:Fock}.}
\label{fig:Hartree}
\end{figure}

The relevant trajectory configurations for the Hartree correction are shown in Fig.\ \ref{fig:Hartree}. There is a one-to-one correspondence between the trajectory configurations for the Fock and Hartree contributions. However, unlike for the Fock contribution, the configurations for the Hartree contribution involve a finite-angle crossing at momenta $\vp_1$ and $\vp_2$, which has two important consequences: First, it leads to an additional difference in the classical actions of the trajectories, resulting in the fast-oscillating factor $e^{i(\vp_1-\vp_2)(\vr_1-\vr_2)/\hbar}$. Such factor also enforces the interaction points $\vr_1$ and $\vr_2$ to remain close together on a scale of the Fermi wavelength. Since the function $K(\vr_1,\vr_2;\omega)$ is built from classical propagators that are smooth 
on the scale of Fermi wavelength, we can identify $\vr_1=\vr_2$ for this function. Second, the spinor structure from the interaction vertices now results in the nontrivial factor
\begin{equation}\eqalign{
 \Sigma_{\rm H}(\vp_1,\vp_2)&=\langle\chi(\vp_1)|\chi(\vp_2)\rangle \langle\chi(\vp_2)|\chi(\vp_1)\rangle\
=\cos^2\left(\frac{\phi_{\vp_1}-\phi_{\vp_2}}{2}\right).
}\end{equation}
This result indeed reflects the chiral nature of the charge carriers, leading to a suppression of backward scattering processes.

Combining everything, the Hartree contribution can be obtained from the Fock contribution by the replacement
\begin{eqnarray}
U^{\rm R}(\vr_1,\vr_2;\omega) &\rightarrow &-\langle U(\vp_1-\vp_2;\omega=0) \Sigma_{\rm H}(\vp_1,\vp_2) \rangle_{\vp_1,\vp_2} 
d_{\rm g}\delta(\vr_1-\vr_2), 
\end{eqnarray}
where the factor $d_{\rm g}$ comes from the existence of a closed trajectory-loop in the configurations of Fig.\ \ref{fig:Hartree}, and with the Fourier-transformed interaction
\begin{equation}
 U^{\rm R}(\vq;\omega)=\int d\vr e^{i\vq\vr/\hbar} U^{\rm R}(\vr;\omega).
\end{equation}
For a short-range potential $U(\vr_1-\vr_2)\propto\delta(\vr_1-\vr_2)$, we then find $\delta G_{\rm AA}^{\rm H}=-(d_{\rm g}/2) \delta G_{\rm AA}^{\rm F}$: The spin and valley degeneracies enhance the Hartree contribution by an extra factor $d_{\rm g} = 4$ compared to the Fock contribution, while chirality reduces it by a factor two \cite{kozikov2010,jobst2012}.

Substituting the explicit expressions for the diffusion propagators, the final result for the interaction correction to the conductivity reads
\begin{eqnarray}
  \label{eq:dGAA-result}
  \delta \sigma_{\rm AA} &=&
  -\frac{d_{\rm g} e^2 \nu D}{\pi \hbar^2} \int d\omega 
  \frac{\partial}{\partial \omega} \left( \omega
  \coth \frac{\omega}{2 T} \right)  \\ && \mbox{} \times
  \mbox{Im}\, \left\lbrace 
  \int \frac{d^2\vq}{(2\pi)^2} {\cal U}^{\rm R}(\vq;\omega) 
  \frac{D q^2 e^{i\omega\tau_{\rm E}}e^{-D\vq^2\tau_{\rm E}}}
     {(D\vq^2-i\omega/\hbar)^3}\right\rbrace,
  \nonumber 
\end{eqnarray}
with the effective interaction kernel
\begin{eqnarray}
  \label{eq:Ueff}
  {\cal U}^{\rm R}(\vq;\omega) &=& U^{\rm R}(\vq,\omega)
  - d_{\rm g} \langle U^{\rm R}(\vp_1-\vp_2;0) 
  \Sigma_{\rm H}(\vp_1,\vp_2) \rangle_{\vp_1,\vp_2}. \nonumber
\end{eqnarray}

\subsection{Coulomb interaction}

The Coulomb interaction, $U_{\rm C}(\vr_1,\vr_2)=e^2/|\vr_1-\vr_2|$ is long-ranged, and screening effects need to be incorporated into the results of the previous section. The treatment of screening requires us to slightly reformulate the results of the previous Section. This discussion closely follows that of \cite{zala2001,gornyi2004,finkelstein1983,altshuler1983,schneider2013}, where the case of Coulomb interactions in a normal metal was considered.

\begin{figure}[t]
\begin{center}
\includegraphics[width=3.6in]{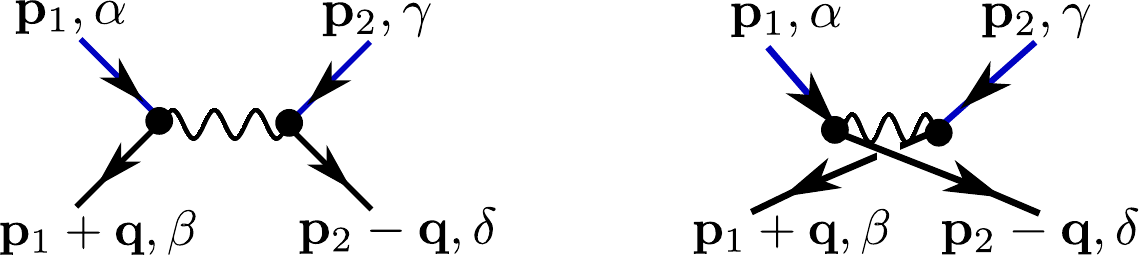}
\end{center}
\caption{Scattering processes of two particles with initial momenta $\vp_1$ ($\vp_2$) and spin+valley $\alpha$ ($\gamma$), and final momenta $\vp_1+\vq$ ($\vp_2-\vq$) and spin+valley $\beta$ ($\delta$) to lowest order interaction.}
\label{fig:interaction0}
\end{figure}

Since all relevant trajectories are paired in the semiclassical analysis, it is sufficient to consider interaction processes in which the transferred momentum $\vq$ is small (eventually after an exchange of the particles). To be specific, we consider the scattering of two particles with initial momenta $\vp_1$ and $\vp_2$, and final momenta $\vp_1+\vq$ and $\vp_2-\vq$, respectively, and assign combined spin/valley indices $\alpha$ and $\gamma$ of the incoming particles, and the indices $\beta$ and $\delta$ to the two outgoing particles, respectively, see Fig.\ \ref{fig:interaction0}. Then the unscreened Coulomb interaction has the matrix elements ${\cal U}_{\alpha\beta\gamma\delta}(\vq) = U_{\rm C}(\vq) \delta_{\alpha,\beta} \delta_{\gamma,\delta} - \langle U_{\rm C}(\vp_1-\vp_2) \Sigma_{\rm H}(\vp_1,\vp_2)\rangle_{\vp_1,\vp_2} \delta_{\alpha,\delta} \delta_{\beta,\gamma}$, where we neglected the small momentum $\vq$ in the argument of $U_{\rm C}$ for the second term and performed the average over the 
directions of $\vp_1$ and $\vp_2$, consistent with the diffusive motion. It is the trace of this averaged matrix element that appears in the effective interaction kernel (\ref{eq:Ueff}) of the perturbative analysis.

The theoretical treatment of screening requires a different structure of the valley and spin indices. The interaction matrix elements are decomposed as
\begin{equation}
 {\cal U}_{\alpha\beta\gamma\delta}(\vq) =\sum_{i=0}^{15} {\cal U}^{(i)}(\vq) \Gamma^{(i)}_{\alpha\beta} \Gamma^{(i)}_{\gamma\delta},
  \label{eq:Uchannels}
\end{equation}
where matrices $\Gamma^{(i)}$ act in spin and valley space, and read \{$s^0\otimes \tau^0,s^0\otimes \tau^k,s^j\otimes \tau^0,s^j\otimes \tau^k$\}, where $s^j$ ($\tau^k$) are Pauli matrices acting in spin (valley) space, with $j,k\in\{x,y,z\}$, and $s^0$ ($\tau^0$) are $2\times 2$-unit matrices. Using the equality $\sum_{i}\Gamma^{(i)}_{\alpha\beta}\Gamma^{(i)}_{\gamma\delta} = 4 \delta_{\alpha\delta} \delta_{\beta\gamma}$ one then finds that the bare Coulomb interaction has
\begin{eqnarray}
  {\cal U}_0^{(i)}(\vq) &=& U_{\rm C}(\vq) \delta_{i,0} 
  -\frac{1}{4}\langle U_{\rm C} (\vp_1-\vp_2) \Sigma_{\rm H} (\vp_1,\vp_2)\rangle_{\vp_1,\vp_2}.
\end{eqnarray}

Screening renormalizes each of the interaction channels ${\cal U}^{(i)}$ in Eq.\ (\ref{eq:Uchannels}) separately. For weak interactions (gas parameter $r_{\rm s} \ll 1$) one may use the random phase approximation, which gives
\begin{equation}
 {\cal U}^{(i)}(\vq;\omega)={\cal U}_0^{(i)}(\vq)-{\cal U}_0^{(i)}(\vq) \Pi(\vq;\omega) {\cal U}^{(i)}(\vq;\omega),
  \label{eq:RPA}
\end{equation}
where $\Pi(\vq;\omega)$ is the polarization operator
\begin{equation}
 \Pi(\vq;\omega)=d_{\rm g}\nu\left[1+\frac{i\omega}{\hbar} P(\vq;\omega)\right],
\end{equation}
$P(\vq;\omega)=(D\vq^2-i\omega/\hbar)^{-1}$ being the diffusion propagator.

The extension of the discussion to stronger interactions requires the use of Fermi liquid theory. Following \cite{zala2001,gornyi2004,finkelstein1983,altshuler1983}, the bare interaction acquires a short-range contribution set by Fermi-liquid constants,
\begin{equation}
  {\cal U}_0^{(0)}(\vq)=U_{\rm C}(\vq)+\frac{1}{d_g \nu} F_0^{\rho}, \quad  {\cal U}_0^{(i\neq 0)}(\vq)=\frac{1}{d_g \nu} F_0^{\sigma},
\end{equation}
where we assume the same Fermi-Liquid constant $F_0^{\sigma}$ in all non-singlet channels. The screened interaction is then still given by Eq.\ (\ref{eq:RPA}).

Applying this procedure to the Altshuler-Aronov correction, we find the same result as the first-order correction (\ref{eq:dGAA-result}), but a different expression for the interaction kernel ${\cal U}^{\rm R}(\vq;\omega)$,
\begin{eqnarray}
 \label{eq:interactionfull}
  {\cal U}^{\rm R}(\vq;\omega) &=&
  \frac{1}{d_{\rm g}\nu} \frac{\hbar D\vq^2-i\omega}{\hbar D\vq^2}
  +
  \frac{F_0^{\sigma}(d_{\rm g}^2-1)(\hbar D\vq^2-i\omega)}{d_{\rm g}\nu [\hbar D\vq^2(1+F_0^{\sigma})-i\omega]}.
\end{eqnarray}
The singlet-channel Fermi liquid constant $F^{\rho}_0$ does not enter into the correction because of the divergence of the Coulomb interaction at small momenta. 

\begin{figure}[t]
\begin{center}
\includegraphics[width=3.5in]{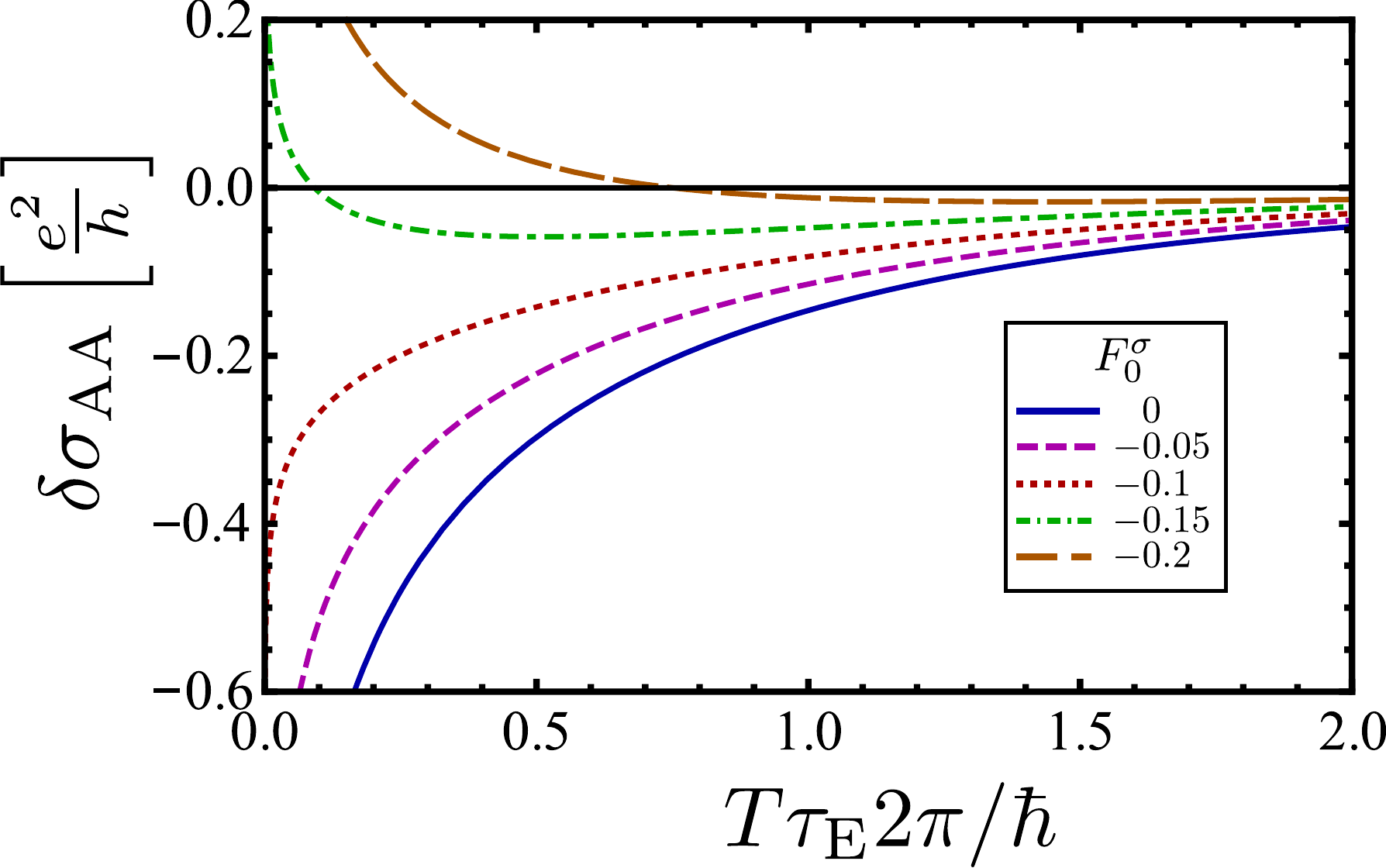}
\end{center}
\caption{Altshuler-Aronov correction as a function of $T\tau_{\rm E} 2\pi/\hbar$. Different curves correspond to different values of the interaction parameter $F_0^{\sigma}$.}
\label{fig:plotAA}
\end{figure}

We plot the Altshuler-Aronov correction for various values of the interaction constant $F_0^{\sigma}$ in Fig.\ \ref{fig:plotAA}. For small Ehrenfest time, one finds
\begin{equation}
 \label{eq:AAlowtE}
 \delta \sigma_{\rm AA}=-\frac{e^2 }{\pi h }\left[1+c\left(1-\frac{\ln(1+F_0^{\sigma})}{F_0^{\sigma}}\right)\right] \ln\frac{\hbar}{2\pi T\tau_{\rm E}},
\end{equation}
with $c=d_{\rm g}^2-1$. Such an expression is well-known from diagrammatic perturbation theory \cite{altshuler1985a}, where  in our case, the Ehrenfest time takes over the role of the elastic scattering time as a short-time cutoff. The graphene-specific physics enters the result in two ways \cite{kozikov2010,jobst2012}: First, the constant $c=d_{\rm g}^2-1$ is 15 for graphene with only smooth disorder, in contrast to $c=3$ for conventional metals without valley degeneracy. Second, chirality affects the interaction constant $F_0^{\sigma}$, as will be explained in more detail below. For small values of $F_0^{\sigma}$, the singlet contribution is dominant in Eq.\ \eref{eq:AAlowtE}, giving rise to a negative correction to the conductance. On the other hand, for graphene, for $F_0^{\sigma}\lesssim -0.12$, the non-singlet channels render the interaction correction positive. 

For large Ehrenfest time, we find an exponential suppression
\begin{equation}
 \delta \sigma_{\rm AA}=-\frac{e^2}{\pi h} e^{-2\pi T\tau_{\rm E}/\hbar}
\end{equation}
where the prefactor of the exponential is determined by the singlet channel only to leading order in $\hbar/ T \tau_{\rm E}$, hence at large Ehrenfest times, $\delta G_{\rm AA}$ is negative and has a universal behavior. 

A striking consequence of this asymptotics is a sign-change of the interaction correction as a function of Ehrenfest time, provided the Fermi-liquid-type interactions in the non-singlet channels are strong enough. For graphene ($c=15$) this sign change already takes place at $F_0^{\sigma}\lesssim -0.12$, in contrast to a conventional metal ($c=3$) where the sign change is observed for $F_0^{\sigma}\lesssim -0.45$. On the other hand, the values for $F_0^{\sigma}$ are typically somewhat smaller in graphene, as can be seen using Thomas-Fermi approximation \cite{kozikov2010,jobst2012}. For conventional metals, one has
\begin{equation}
 F_0^{\sigma}=-\nu\left\langle \frac{2\pi e^2_{\star}}{|\vp_1-\vp_2|+\kappa} \right\rangle_{\vp_1,\vp_2},
\end{equation}
where $e_{\star}$ is the charge screened by the substrate \cite{kozikov2010,jobst2012}, and $\kappa=2\times2\pi\nu e^2_{\star}$ is the inverse screening length resulting from the metal electrons (a factor $2$ accounts for spin). We then find
\begin{equation}
 F_0^{\sigma}
      =-\nu\int_0^{\pi} \frac{d\theta}{\pi} \frac{2\pi e^2_{\star}}{2 k_F \sin\frac{\theta}{2}+4\pi\nu e^2_{\star}}
      =-\int_0^{\pi} \frac{d\theta}{\pi} \frac{\alpha}{2 \sin\frac{\theta}{2}+2\alpha},
\end{equation}
where $\theta$ is the angle between the directions of momenta $\vp_1$ and $\vp_2$. We further used $k_{\rm F}=2\pi \nu \hbar v_{\rm F}$, as well as the ``effective fine structure constant'' $\alpha={e_{\star}^2}/{\hbar v_{\rm F}}$. The gas  parameter $r_{\rm s}$ is related to $\alpha$ as $r_{\rm s}=\sqrt{2}{\alpha}$. For a value $r_s\approx 1$ we obtain $F_0^{\sigma}\approx-0.28$ as a typical size for the Fermi liquid parameter.

For graphene, this calculation needs to be modified in two respects. First, the inverse screening length is twice larger, due to the valley degree of freedom. Second, chirality contributes the additional factor $\Sigma_{\rm H}(\vp_1,\vp_2)= \cos^2(\theta/2)$. Both effects reduce the interaction in the non-singlet channel,
\begin{equation}\eqalign{
 F_0^{\sigma}=-\int_0^{\pi} \frac{d\theta}{\pi} \frac{\alpha \cos^2\frac{\theta}{2}}{2 \sin\frac{\theta}{2}+4\alpha},
}\end{equation}
so that now for $r_{\rm s}\approx 1$ we find $F_0^{\sigma}\approx-0.1$, which is close to the transition point for a sign-change as function of Ehrenfest time.
The measurements of \cite{kozikov2010,jouault2011,jobst2012} report $F_0^{\sigma}$ in a range between $-0.05$ and $-0.15$. However, we note that our theory requires graphene with a smooth disorder potential, but no other perturbations, as a necessary condition for the value $c=15$ in Eq.\ (\ref{eq:AAlowtE}), since there are $16$ diffusion channels. Trigonal warping or ripples, while not invalidating the semiclassical analysis, reduce the number of diffusive channels to $8$, resulting in a prefactor $c=7$. In case of strong intervalley scattering, only four diffusion modes are present resulting in a prefactor $c=3$ (see Ref.\ \cite{kechedzhi2008,kharitonov2008}) --- although in that case the conditions for the semiclassical analysis are no longer valid. The aforementioned experiments on interaction corrections report to be in a regime where $c=3$ or $c=7$.

\section{Dephasing}
\label{sec:Dephasing}
We now turn to the second type of interaction correction, responsible for dephasing. The way of calculating dephasing in this section follows that of Ref.\ \cite{altshuler1982b}. An alternative discussion based on perturbation theory, similar to that of the previous Section, is given in the appendix.

Because of the interactions, the electrons are subject to a time-dependent potential $V(\vr,t)$. This potential affects the phase that electrons accumulate while propagating through the sample. These phase fluctuations can be included into the classical action ${\cal S}_{\alpha}$ of the trajectory $\alpha$ as it appears in the semiclassical expression (\ref{eq:SemiGreen}) for the Green function by the substitution
\begin{equation}
  {\cal S}_{\alpha} \to {\cal S}_{\alpha} + \delta {\cal S}_{\alpha}(t),
\end{equation}
where the correction $\delta {\cal S}_{\alpha}(t)$ depends on the time $t$ at which the electron exits the sample. The shift reads \cite{altshuler1982b,altland2007}
\begin{equation}
  \delta {\cal S}_{\alpha}(t) = \int_{t-\tau_{\alpha}}^{t} dt' V[\vr_{\alpha}(t'),t'],
\end{equation}
where $\tau_{\alpha}$ is the duration of the trajectory $\alpha$.

Such a shift of the classical actions does not affect the Drude conductivity, because the actions from the ``retarded'' and ``advanced'' trajectories cancel. It does, however, affect the weak antilocalization correction. Equation (\ref{eq:dGWAL}) acquires an additional factor $e^{i(\delta {\cal S}_{\alpha}(t) - \delta {\cal S}_{\beta}(t))}$ which, when averaged over the time $t$, reduces the contribution from the trajectory pair $\alpha$, $\beta$ by a factor
\begin{equation}
  \left\langle 
  e^{i(\delta {\cal S}_{\alpha}(t) - \delta {\cal S}_{\beta}(t))/\hbar}
  \right\rangle_t =
  e^{-(1/2 \hbar^2) \langle (\delta {\cal S}_{\alpha}(t) - \delta {\cal S}_{\beta}(t))^2 \rangle_t}.
  \label{eq:Scorr}
\end{equation}
The time average can be calculated using the quantum fluctuation-dissipation theorem,
\begin{eqnarray}
  \langle V(\vr,t) V(\vr',t') \rangle &=&
  \int \frac{d^2 \vq d\omega}{(2 \pi)^3}
  \frac{\omega}{2 T \sinh^2(\omega/2 T)}
  e^{i \vq \cdot (\vr-\vr') - i \omega (t-t')}
  \mbox{Im}\, U^{\rm R}(\vq,\omega), \nonumber
\end{eqnarray}
which gives
\begin{eqnarray}
  &&\frac{1}{2}
  \langle ((\delta {\cal S}_{\alpha}(t) - \delta {\cal S}_{\beta}(t))^2 \rangle_t 
  \nonumber\\ &=&
  \int_0^{\tau_{\alpha}} dt_1 \int_0^{t_1} dt_2
  \int \frac{d^2 \vq d\omega}{(2 \pi)^3}
  \frac{\omega}{2 T \sinh^2(\omega/2 T)}
  \mbox{Im}\, U^{\rm R}(\vq,\omega)
  \nonumber \\ && \mbox{} \times
  \mbox{Re}\, e^{-i \omega(t_1-t_2)}
  (e^{i \vq \cdot \vr_{\alpha}(t_1)} - e^{i \vq \cdot \vr_{\beta}(t_1)})
  (e^{-i \vq \cdot \vr_{\alpha}(t_2)} - e^{-i \vq \cdot \vr_{\beta}(t_2)})
  \label{eq:deph}
\end{eqnarray}
(Note that $\tau_{\alpha} = \tau_{\beta}$ for the trajectory pairs that contribute to weak antilocalization.)
One immediately concludes that for the trajectories $\alpha$ and $\beta$ that contribute to the weak antilocalization correction $\delta G_{\rm WAL}$ only points $\vr_{\alpha}$ or $\vr_{\beta}$ in the loop or encounter segments of Fig.\ \ref{fig:WL} contribute to $\langle ((\delta {\cal S}_{\alpha}(t) - \delta {\cal S}_{\beta}(t))^2 \rangle_t$. 

To find an explicit expression for the dephasing correction in the limit of weak dephasing, we expand the correction factor (\ref{eq:Scorr}) to lowest order in the interaction $U^{\rm R}(\vq,\omega)$ and calculate the leading interaction correction $\delta G_{\rm deph}$ to the weak antilocalization correction $\delta G_{\rm WAL}$. We consider contributions from positions $\vr_{\alpha,\beta}(t_1)$ and $\vr_{\alpha,\beta}(t_2)$ in the loop and encounter regions separately.

\begin{figure}[t]
\begin{center}
\includegraphics[width=4.3in]{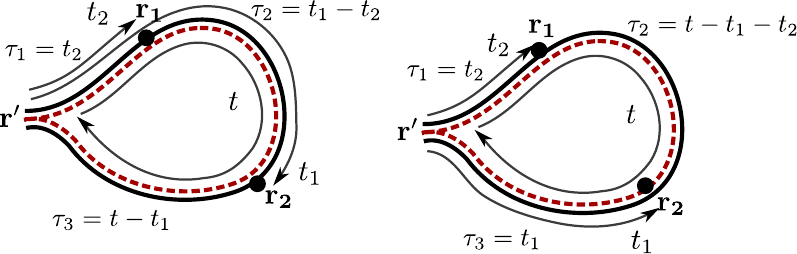}
\end{center}
\caption{Dephasing in the loop segment: for the calculation of the lowest-order interaction correction the propagation around the loop is split into three segments $\vr'\rightarrow\vr_1$ (duration $\tau_1$), $\vr_1\rightarrow\vr_2$ (duration $\tau_2$), and $\vr_2\rightarrow\vr'$ (duration $\tau_3$). For our figure, the trajectories $\alpha$ (solid) and $\beta$ (dashed) are travelled in clockwise and counterclockwise direction, respectively. The left diagram represents the contribution from the term proportional to $e^{i \vq \cdot \vr_{\alpha}(t_1)-i \vq \cdot \vr_{\alpha}(t_2)}$ in Eq.\ (\ref{eq:deph}); the right diagram represents the contribution from the term proportional  $e^{i \vq \cdot \vr_{\beta}(t_1)-i \vq \cdot \vr_{\alpha}(t_2)}$. The remaining two contributions, from terms proportional to $e^{i \vq \cdot \vr_{\alpha}(t_1)-i \vq \cdot \vr_{\beta}(t_2)}$ and $e^{i \vq \cdot \vr_{\beta}(t_1)-i \vq \cdot \vr_{\beta}(t_2)}$ are not shown. The figures have been drawn for the case that $0 
< 
t_2 < \min(t_1,t-t_1)$, which is the domain of integration in Eq.\ (\ref{eq:deltaP}).}
\label{fig:deph}
\end{figure}

The calculation for the dephasing in the loop segment is very similar to the one carried out in standard diagrammatic perturbation theory. The discussion below closely follows that of \cite{marquardt2007}. With both positions $\vr_{\alpha,\beta}(t_{1,2})$ in the loop region, see Fig.\ \ref{fig:deph}, we find that dephasing in the loop segment leads to the replacement $P(\vr',\vr';t) \to P(\vr',\vr';t) + \delta P(\vr',\vr';t)$ for the loop propagator in Eq.\ \eref{eq:WALexpr}, with
\begin{eqnarray}
 \label{eq:deltaP}
  \delta P(\vr',\vr';t) &=&-\frac{4}{\hbar^2}
  \int\frac{d\vq}{(2\pi)^2} \int\frac{d\omega}{2\pi}
  \frac{\omega\, \Im U^{R}(\vq;\omega)}{2 T \sinh^2(\omega/2 T)}
  \nonumber\\ && \mbox{} \times
  \int_{0}^{t} dt_1
  \int_{0}^{\min(t_1,t-t_1)} dt_2  \cos[\omega (t_1-t_2)/\hbar]
  \nonumber\\ && \mbox{} \times  
  \left[{\cal P}_{\vq}(t_2,t_1-t_2,t-t_1)-{\cal P}_{\vq}(t_2,t-t_1-t_2,t_1)\right],
\end{eqnarray}
where
\begin{eqnarray}
 {\cal P}_{\vq}(\tau_1,\tau_2,\tau_3) &=&
  \int d\vr_1 d\vr_2 \cos[\vq \cdot (\vr_1-\vr_2)] \nonumber\\ &&\times 
  P(\vr',\vr_2,\tau_3) 
  P(\vr_2,\vr_1,\tau_2) P(\vr_1,\vr',\tau_1).
\end{eqnarray}
Inserting the diffusion propagator $P(\vq,\tau)=e^{-D q^2\tau}$, one finds
\begin{equation}
 \label{eq:calP}
 {\cal P}_{\vq}(\tau_1,\tau_2,\tau_3)=P(\vr',\vr';t)
  e^{-Dq^2\tau_2(1-\tau_2/t)},
\end{equation}
with $t=\tau_1+\tau_2+\tau_3$.
After insertion of the interaction \eref{eq:interactionfull} and evaluation of the integrals over time, frequency, and momentum in Eq.\ \eref{eq:deltaP} (see Appendix \ref{sec:appdeph}), one obtains
\begin{eqnarray}
 \label{eq:prop_loop}
 \frac{\delta P(\vr',\vr';t)}{P(\vr',\vr';t)} = - \frac{\alpha tT}{\hbar g_0}\ln\frac{tT}{\hbar},
\end{eqnarray}
with the dimensionless conductance $g_0=2\pi\hbar d_{\rm g}\nu D$, and the constant
\begin{equation}
 \alpha=1+(d_{\rm g}^2-1)\frac{\left(F_0^{\sigma}\right)^2}{\left(1+F_0^{\sigma}\right)\left(2+F_0^{\sigma}\right)}.
\end{equation}

This result signifies that at large times the loop propagator gets suppressed by interactions. Here we calculated the leading order correction, describing the onset of an exponential suppression. (In fact, in two dimensions the decay is not purely exponential \cite{marquardt2007}, but contains an additional logarithm $e^{-at\ln t}$, as can be seen from Eq.\ \eref{eq:prop_loop}). We estimate the dephasing rate as the time when the leading correction becomes unity, {\em i.e.},
\begin{equation}
 \frac{1}{\tau_{\varphi}^{\rm loop}}\simeq\frac{\alpha T}{g_0\hbar}\ln \frac{g_0}{\alpha}.
\end{equation}
The leading logarithmic dependence in this expression agrees with that obtained in \cite{narozhny2002} for a standard two-dimensional electron gas ($d_{\rm g} = 2$).

We now turn to the encounter region, where dephasing leads to an additional suppression of weak antilocalization, if the typical time for the encounter passage, the Ehrenfest time $\tau_{\rm E}$, is sufficiently long. As discussed before, dephasing is ineffective, as long as the trajectories coincide. Within the encounter region, the trajectories $\alpha$ and $\beta$ are separated by a small distance, which does not exceed the classical correlation scale $L_{\rm c}$. Dephasing then only plays a role for interaction that transfers a momentum larger than inverse mean free path, and therefore can resolve such small distance \cite{altland2007,petitjean2007,whitney2008}. On the other hand, for low temperatures $T \tau_{\rm tr} \ll 1$ one has $\omega \tau_{\rm tr} \ll 1$. In this limit, the imaginary part of the screened interaction reads \cite{zala2001}
\begin{equation}
 \label{eq:ImU}
  \Im U^{R}(\vq;\omega)= -
  \frac{\beta \omega}{q d_{\rm g} \hbar \nu v_{\rm F}},
\end{equation}
where $\nu=k_{\rm F}/2\pi\hbar\vF$ is the density of states and we 
abbreviated 
\begin{equation}
  \beta = 
  1+(d_g^2-1)\frac{\left(F_0^{\sigma}\right)^2}{\left(1+F_0^{\sigma}\right)^2}.
\end{equation}
Note that $\Im\, U^{\rm R}(\vq;\omega)$ is proportional to $q^{-1}$, which is different from the dependence $\Im\, U^{R}(\vq;\omega) \propto q^{-2}$ of the diffusive limit. This difference will result a different $T$-dependence of the dephasing rate in comparison to the loop contribution \cite{altland2007}.

We proceed by the integration over $\omega$, which can be done explicitly using the known $\omega$ dependence of $\Im U^{R}(\vq;\omega)$ \cite{marquardt2007},
\begin{eqnarray}
 \label{eq:omegaint}
  \int \frac{d\omega}{2\pi} \frac{\omega^2 e^{-i\omega (t_1-t_2)/\hbar}}{2 T \sinh^2 (\omega/2 T)} =
  2\pi T^2 w[\pi T (t_1-t_2)/\hbar].
\end{eqnarray}
Here the function $w(x)=(x\coth x -1)/\sinh^2 x$ is peaked around $x=0$, normalized $\int_{-\infty}^{\infty} dx w(x)=1$, and $w(0)=1/3$. Hence, the times $t_1$ and $t_2$ need to be close together on the scale of inverse temperature. On the other hand, dephasing sets in on times much larger than $T^{-1}$, as we will show below. In the following, we therefore may assume that $|t_1-t_2|\ll\tau_{\rm E}$, when we consider encounters that are long enough to be affected by dephasing. (In fact, the calculation below will show that the main contribution stems from time differences $|t_1 - t_2|$ much smaller than the elastic mean free time.) In particular this allows us to consider the effect of interaction during the first and second passage through the encounter separately, since they are separated by a loop of long duration. The same observation also allows us to neglect contributions where $t_1$ is in the encounter, whereas $t_2$ is in the loop or vice versa. We therefore focus on the first passage through the 
encounter region, where the trajectories $\alpha$ and $\beta$ are separated by a distance $\vd(t)=\vr_{\beta}(t)-\vr_{\alpha}(t)$, with the magnitude 
\begin{equation}
 \label{eq:dt}
 d(t) \simeq \lambda_F e^{\lambda t}, 
\end{equation}
where $t$ is varying form $0$ to $\tau_{\rm E}$. We can use this to rewrite the last two factors of Eq.\ \eref{eq:deph} as
\begin{eqnarray}
  \lefteqn{
  (e^{i \vq \cdot \vr_{\alpha}(t_1)} - e^{i \vq \cdot \vr_{\beta}(t_1)})
  (e^{-i \vq \cdot \vr_{\alpha}(t_2)} - e^{-i \vq \cdot \vr_{\beta}(t_2)})}
  \nonumber \\ &=&
  4 e^{i \vq \cdot [\bar \vr(t_1) - \bar \vr(t_2)]}
  \sin[\vq \cdot \vd(t_1)/2] \sin[\vq \cdot \vd(t_2)/2],~~~~
\end{eqnarray}
where $\bar \vr(t) = [\vr_{\alpha}(t) + \vr_{\beta}(t)]/2$ represents a trajectory intermediate between $\alpha$ and $\beta$.

\begin{figure}[t]
\begin{center}
\includegraphics[width=2.2in]{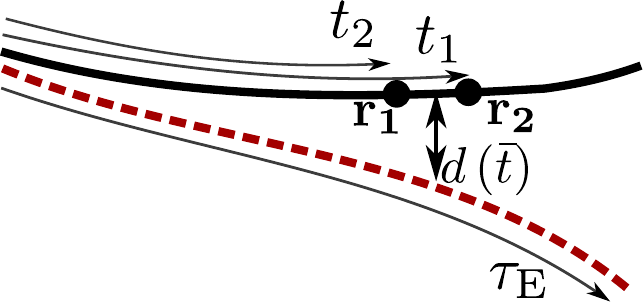}
\end{center}
\caption{Dephasing in the encounter segment: For the calculation, the encounter is split into three segments $\vr'\rightarrow\vr_1$ (duration $\tau_1=t_2$), $\vr_1\rightarrow\vr_2$ (duration $\tau_2=t_1-t_2$), and $\vr_2\rightarrow\vr'$ (duration $\tau_3=\tau_{\rm E}-t_2$). Only configurations in which $|t_1-t_2| \ll \tau_{\rm E}$ contribute to the interaction correction $\delta G_{\rm deph}$. In the middle segment, the distance between the trajectories $\alpha$ and $\beta$ is $d(\bar{t})=\lambda_{\rm F} e^{\lambda \bar{t}}$, where $\bar{t}=(t_1+t_2)/2$. The figure has been drawn for the case $0 < t_2 < t_1$, which is the domain of integration in Eq.\ (\ref{eq:deltaP_enc})}
\label{fig:deph_enc}
\end{figure}

After performing the average over disorder configurations, we find that inclusion of the leading-order dephasing correction amounts to the replacement $P(\vr',\vr;\tau_{\rm E}) \to P(\vr',\vr;\tau_{\rm E}) + 2\delta P(\vr',\vr;\tau_{\rm E})$ in Eq.\ (\ref{eq:WALexpr}), where the factor two accounts for the two passages through the encounter region, with
\begin{eqnarray}
 \label{eq:deltaP_enc}
 \fl \delta P(\vr',\vr;\tau_{\rm E}) =
  -\frac{8\pi T^2 \beta}{\hbar^3 v_{\rm F} d_{\rm g} \nu}\int\frac{d^2\vq}{(2\pi)^2} \frac{1}{ q} \int_0^{\tau_{\rm E}} dt_1 \int_0^{t_1} dt_2
  w(\pi T (t_1-t_2)/\hbar)
  \nonumber \\  \fl  \qquad \qquad \qquad \mbox{} \times
  {\cal P}^{\rm enc}_{\vq}(t_1,t_2,\tau_{\rm E}-t_1-t_2)
  \sin[\vq \cdot \vd(t_1)/2] \sin[\vq \cdot \vd(t_2)/2]
\end{eqnarray}
and
\begin{eqnarray}
  \fl {\cal P}_{\vq}^{\rm enc}(t_1,t_2,\tau_{\rm E} - t_1 - t_2) = 
  \int d\vr_1 d\vr_2 \cos[\vq \cdot (\vr_1-\vr_2)] \nonumber \\
  \fl \qquad \qquad \qquad \qquad \qquad \mbox{} \mbox{} \times P(\vr',\vr_2,\tau_{\rm E} - t_1) P(\vr_2,\vr_1,t_1-t_2) P(\vr_1,\vr,t_2), 
\end{eqnarray}
see Fig.\ \ref{fig:deph_enc}.
Because of the smallness of $|t_1-t_2|$, the propagator $P(\vr_2,\vr_1,t_1-t_2)$ is the ballistic propagator, whereas the propagators $P(\vr',\vr_2,\tau_{\rm E} - t_1)$ and $P(\vr_1,\vr,t_2)$ can be taken in the diffusion approximation. We change the integration variables to the mean time $\bar t$ and the difference time $t = t_1-t_2$. Again using the smallness of $|t_1-t_2|$, we replace $\vd(t_1)$ and $\vd(t_2)$ by $\vd(\bar t)$. We neglect correlations between $\vd(\bar t)$ and the direction of the velocity at time $\bar t$. Using $P(\vr_1,\vr,t_2) \simeq P(\vr_2,\vr,t_1)$, again because of the smallness of $|t_1-t_2|$, we find 
\begin{equation}
  {\cal P}_{\vq}^{\rm enc}(t_1,t_2,\tau_{\rm E} - t_1 - t_2) \simeq 
  P(\vr',\vr;\tau_{\rm E}) J_0(v_{\rm F} q |t|),
\end{equation}
where we inserted the Fourier transform of the ballistic propagator. Since the integration over $t$ converges for $|t| \sim 1/v_{\rm F} q$, the argument of the function $w$ may be set to zero in Eq.\ (\ref{eq:deltaP_enc}). Finally, the angular average over the direction of $\vd(\bar t)$ gives a factor $1 - J_0(q d(\bar t))$, so that we find
\begin{eqnarray}
 \label{eq:a1}
   \delta P(\vr',\vr;\tau_{\rm E}) &=&
  -\frac{4 \pi T^2 \beta}{3 \hbar^3 v_{\rm F}^2 d_{\rm g} \nu}
  P(\vr',\vr;\tau_{\rm E})
  \int\frac{d^2\vq}{(2\pi)^2} \frac{1}{q^2} 
  \nonumber \\ && \mbox{} \times
  \int_0^{\tau_{\rm E}} d\bar t
  \left[1-J_0\left(qd(\bar{t})\right)\right],
\end{eqnarray}  

We cut off the logarithmic divergence of the $\vq$ integration at large $q$ at $\lambda_{\rm F}^{-1}$, which gives
\begin{eqnarray}
 \label{eq:a2}
   \delta P(\vr',\vr;\tau_{\rm E}) &=&
  -\frac{2  T^2 \beta}{3 \hbar^3 v_{\rm F}^2 d_{\rm g} \nu}
  P(\vr',\vr;\tau_{\rm E})
  \int_0^{\tau_{\rm E}} d\bar t
  \ln \frac{d(\bar t)}{\lambda_{\rm F}}.
  \nonumber \\
\end{eqnarray}  
The remaining time-integration is easily evaluated with the help of Eqs.\ \eref{eq:dt}, \eref{eq:tauE}, and we obtain
\begin{equation}
 \int_0^{\tau_{\rm E}} d\bar{t}  \ln\frac{d(\bar{t})}{\lambda_{F}}=\frac{\tau_{\rm E}}{2} \ln\frac{L_{\rm c}}{\lambda_{\rm F}}.
\end{equation}
Hence, our the final result reads
\begin{equation}
\label{eq:Penc}
  \frac{\delta P(\vr',\vr;\tau_{\rm E})}{P(\vr',\vr,\tau_{\rm E})}
  = - 
  \frac{ T^2 \beta \tau_{\rm E}}{3 \hbar^3 v_{\rm F}^2 d_{\rm g} \nu}
  \ln\frac{L_{\rm c}}{\lambda_{\rm F}}.
\end{equation}

One may identify the right-hand side of Eq.\ (\ref{eq:Penc}) with $\tau_{\rm E}/\tau_{\varphi}^{\rm enc}$, where $\tau_{\varphi}^{\rm enc}$ is an effective dephasing time for the encounter region. With this identification, Eq.\ \eref{eq:Penc} describes the onset of an exponential suppression of the weak localization $\propto e^{-2 \tau_{\rm E}/\tau_{\varphi}^{\rm enc}}$ at large Ehrenfest times. Note that the time $\tau_{\rm \varphi}^{\rm enc}$ is {\em twice} the dephasing time $\tau_{\varphi}^{\rm ball}$ that one finds from dephasing in the loop region in the ballistic regime \cite{narozhny2002}, consistent with the theory of \cite{altland2007}. [To compare with \cite{narozhny2002}, take the energy-dependent dephasing time $\tau_{\varphi}(\varepsilon)$ from Eqs.\ (18) and (19a) of Ref.\ \cite{narozhny2002} in the limit $T\tau\ll\hbar$ and calculate $(\tau_{\varphi}^{\rm ball})^{-1}=\int d\varepsilon \left(-{\partial f}/{\partial \varepsilon}\right) \tau_{\varphi}(\varepsilon)^{-
1}$ with, for 
a conventional metal, $d_{\rm g}=2$. The low-momentum cut-off in \cite{narozhny2002} is the inverse mean free path, whereas it is the classical correlation length $L_{\rm c}$ in our case. The two lengths need not be equal, see Eq.\ (\ref{eq:mfp}).]

\section{Conclusion}
\label{sec:conclusion}

In this article we have presented a trajectory-based semiclassical theory of the quantum corrections to transport in graphene in the presence of a random potential that is smooth on the scale of the Fermi wavelength. A prominent role is played by the Ehrenfest time, which serves as a short-time threshold for the appearance of quantum interference effects. The Ehrenfest time also plays an important role for electrons in a conventional two-dimensional electron gas (with quadratic dispersion) if they are subject to a smooth random potential. 

Compared to the conventional case, charge carriers in graphene have an additional pseudospin degree of freedom and they have an additional valley degeneracy, which leads to a few subtle modifications of the quantum corrections with respect to the conventional case. The pseudospin vector always points along the direction of motion, reflecting the chiral nature of the charge carriers in graphene. The evolution of the pseudospin along the trajectory is associated with a Berry phase of the spin transport, that additionally enters the semiclassical Green function. This Berry phase is responsible for a sign change in the weak localization correction, giving antilocalization behavior. The presence of a finite Ehrenfest time reduces the magnitude of this correction, but with a multiplicative factor that is the same for weak localization and weak antilocalization. We also considered the suppression of weak (anti)localization from dephasing at finite temperatures, and identified there, too, the role of the Ehrenfest 
time.

For the interaction correction there are two important differences with the case of the conventional two-dimensional electron gas: The Hartree-type processes (or, more precisely, interaction non-singlet channels) contain an additional angular dependence, as a result of chirality. Moreover, importance of screening is changed, because of the presence of the valley degeneracy. A finite Ehrenfest time suppresses the Altshuler-Aronov correction, in a similar way as for conventional metals, but unlike for weak (anti)localization the suppression is not simply a multiplicative factor. Interestingly, the interaction correction may undergo a sign change as a function of Ehrenfest time, for sufficiently strong interaction in the non-singlet channels. For graphene, the interaction strength at which this sign change takes place is smaller than in conventional electron gases, which may place it within experimental reach, as discussed in Sec.\ \ref{sec:AA}. 

\ack
We gratefully acknowledge discussions with Falko Pientka and Peter Silvestrov.
This work is supported by the Alexander von Humboldt Foundation in the framework of the Alexander von Humboldt Professorship, endowed by the Federal Ministry of Education and Research and by the German Research Foundation (DFG) in the framework of the Priority Program 1459 ``Graphene''.

\appendix

\section{Lyapunov coefficient}
\label{sec:applyapunov}

Here we calculate the Lyapunov coefficient $\lambda$ for Gaussian-correlated disorder potential $V(\vr)$, specified by Eq.\ \eref{eq:Vcorr}. Our result agrees with that of Aleiner and Larkin \cite{aleiner1996}, using a different method.

We consider two trajectories that are initially close in phase space, and investigate their divergence as they evolve in time. We use $\Delta r_{\perp}$ and $\Delta p_{\perp}$ to denote the position and momentum differences in the direction perpendicular to the propagation direction. From the classical equation of motion,
\begin{equation}
  \dot{\vr} = v_{\rm F} \ve_{\vp},\ \
  \dot{\vp} = - \vnabla V,
\end{equation}
with $\ve_{\vp}$ the unit vector in the direction of the momentum $\vp$, we find that the differences $\Delta r_{\perp}$ and $\Delta p_{\perp}$ evolve in time as
\begin{equation}\eqalign{
  \frac{\partial \Delta r_{\perp}}{\partial t} =
  \vF \frac{\Delta p_{\perp}}{\pF}, \quad 
  \frac{\partial \Delta p_{\perp}}{\partial t}
  =
  -\frac{\partial^2 V(t)}{\partial r_{\perp}^2} \Delta r_{\perp},
}\end{equation}
where $V(t)$ is shorthand notation for $V(\vr(t))$. Upon integrating the evolution equations for an infinitesimal time interval $\delta t$ the solution may be cast in the form of a transfer matrix equation, which we write as
\begin{equation}\eqalign{
 \vect{\frac{\Delta r_{\perp}(t+\delta t)}{\xi}}{\frac{\Delta p_{\perp}(t+\delta t)}{z \pF}}&={\cal M}(t+\delta t,t)\vect{\frac{\Delta r_{\perp}(t)}{\xi}}{\frac{\Delta p_{\perp}(t)}{z \pF}},
}\end{equation}
where $z^2 = \sqrt{K_0}/k_{\rm F} \xi \ll 1$ and the transfer matrix ${\cal M}(t,t+\delta t)$ reads
\begin{equation}\eqalign{
  {\cal M}(t+\delta t,t) = e^{z H(t) \delta t},\ \
  H(t) = \matr{0}{\vF/\xi}{f(t)}{0},
}\end{equation}
with
\begin{equation}
  f(t) = - \frac{\xi}{z^2 p_{\rm F}}
  \frac{\partial V(t)}{\partial r_{\perp}^2}
\end{equation}
a stochastic function that contains all information on the random potential. 
The function $f$ has zero mean, and its fluctuations in a time interval $\Delta t$ long in comparison to the correlation time $t_{\xi} = \xi/v_{\rm F}$ are
\begin{equation}\eqalign{
 \left\langle \int_0^{\Delta t} dt dt'f(t) f(t')\right\rangle&=\frac{3}{\sqrt{2\pi}} \frac{\vF \Delta t}{\xi}.
}\end{equation}
(The condition $\Delta t \gg t_{\xi}$ is consistent with the smallness of the parameter $z$.)

Sofar we have calculated the transfer matrix for an infinitesimal time interval $\delta t$. The result can be easily extended to calculate the transfer matrix for time intervals of arbitrary duration, via successive multiplication of transfer matrices valid for the infinitesimal segments. This results in a stochastic evolution of the transfer matrix, which can be analyzed using an explicit parameterization of the transfer matrix,
\begin{equation}
  {\cal M}(t+\Delta t,t)
  = e^{i \varphi \sigma_2} e^{l \sigma_3} e^{i \phi \sigma_2},
\end{equation}
where $\sigma_2$ and $\sigma_3$ are the Pauli matrices. The exponential divergence of the trajectories follows from the radial parameter $l$, 
\begin{equation}
  \lambda = \lim_{\Delta t \to \infty} \frac{l}{\Delta t}.
  \label{eq:lyapunov_def}
\end{equation}
For the calculation of $l$ it is sufficient to consider the matrix product ${\cal M}^T {\cal M}$, which has eigenvalues $e^{\pm 2 l}$ and no longer depends on the angular variable $\varphi$. The time-evolution of the remaining parameters $l$ and $\phi$ is given by a Langevin-type process which, for large $l$, reads
\begin{equation}\eqalign{
  \delta l= & \frac{z}{2}\frac{\vF \delta t}{\xi}\sin 2\phi +
  \frac{z}{2} \sin 2 \phi\int_0^{\delta t} dt'f(t') \nonumber\\
          &-\frac{z^2}{2}\cos2\phi \sin^2 \phi  \int_0^{\delta t} dt' dt'' f(t') f(t''),\nonumber\\
 \delta \cot \phi=& z \int_0^{\delta t} dt' f(t')-z\frac{\vF \delta t}{\xi}\cot ^2 \phi,
}\end{equation}
where terms of higher order than $\delta t$ are neglected.

It is helpful to introduce the variable $y$ via 
\begin{equation}
  \label{eq:ydef}
  y= (2 \pi)^{1/6} z^{-1/3} (2/3)^{1/3} \cot \phi.
\end{equation}
After averaging over fluctuations of $f$, we find that mean and variance of the change $\delta y$ in an infinitesimal time interval $\delta t$ read
\begin{equation}\eqalign{
 \label{eq:deltay}
 \langle\delta y\rangle&=-
  (3/2)^{1/3} (2 \pi)^{-1/6} z^{4/3}
  y^2\frac{\vF \delta t}{\xi}, \nonumber\\
\langle(\delta y)^2\rangle&=
  2^{2/3} 3^{1/3} (2 \pi)^{-1/6} z^{4/3}
  \frac{\vF \delta t}{\xi}.
}\end{equation}
The parameter $y$ acquires a stationary probability distribution $P_{\rm s}(y)$, which satisfies the equation \cite{vankampen2007}
\begin{equation}
 \frac{1}{2}\frac{\langle (\delta y) ^2\rangle}{\delta t} \frac{\partial P_{\rm s}}{\partial y}-\frac{\langle \delta y \rangle}{\delta t} {P_{\rm s}}= c',
\end{equation}
where $c'$ is a numerical constant. Using Eq. \eref{eq:deltay}, we obtain
\begin{equation}
 \frac{\partial P_{\rm s}}{\partial y}+y^2 P_{\rm s}= c'.
\end{equation}
The only normalized solution of this equation occurs for $1/c' = 3^{-5/6} 2^{1/3} \Gamma(1/6) \sqrt{\pi} \approx 4.976$ and reads
\begin{equation}
\label{eq:Ps}
 P_{\rm s}(y)= c' \int_{-\infty}^y dy' e^{(y'^3-y^3)/3}.
\end{equation}
Keeping leading terms in the parameter $z$ only, we find that the average
\begin{equation}
 \langle \delta l \rangle= \left(\frac{3 z^4}{2\sqrt{2\pi}}\right)^{1/3} y \frac{v_{\rm F} \delta t}{\xi}.
\end{equation}
Since the angular variable $y$ evolves statistically independent from the radial variable $l$ for large $l$, we may average $y$ with the help of the stationary distribution \eref{eq:Ps}, for which we find
\begin{equation}
 \beta=\int_{-\infty}^{\infty} dy y P_{\rm s} (y) 
  = \frac{(3/2)^{1/3} \sqrt{\pi}}{\Gamma(1/6)} \approx 0.365.
\end{equation}
From the definition (\ref{eq:lyapunov_def}) we then obtain the Lyapunov coefficient
\begin{equation}
 \lambda= \beta \frac{v_F}{\xi}\left(\frac{3 K_0}{2\sqrt{2\pi}(k_F \xi)^2}\right)^{1/3} = \frac{\beta}{\tau_{\rm tr}} \left( \frac{l_{\rm tr} \sqrt{3}}{\xi} \right)^{2/3},
\end{equation}
where, in the second equality, we inserted the transport mean free time and the mean free path of Eq.\ (\ref{eq:mfp}).
This result agrees with the Lyapunov exponent calculated by Aleiner and Larkin \cite{aleiner1996}. (One has to identify the short-length cut-off $a$ of Ref.\ \cite{aleiner1996} with $\xi/\sqrt{3}$, see the text below Eq.\ (A3) of \cite{aleiner1996}.)

\section{Weak antilocalization}
\label{sec:appWAL}
Here we present some details of the calculation to the weak antilocalization. We start from Eq.\ \eref{eq:WALexpr} and insert the diffusion propagators Eq.\ \eref{eq:diffpropagator}. After performing the spatial integrals, we find
\begin{eqnarray}
  \delta G_{\rm WAL} &=&
  \frac{e^2 d_{\rm g}}{2\pi\hbar}\frac{64}{\pi^4}\sum_{n=1,\, n\,\mathrm{odd}}^{\infty} \sum_{l=1}^{\infty} \sum_{k=0}^{\infty} 
     \frac{e^{-(l^2+k^2/r^2 )\frac{\tau_{\rm tr}}{\tau_{\rm D}}}}{l^2+k^2/r^2}
  \frac{l^2 e^{-n^2 \frac{\tau_{\rm E}}{\tau_{\rm D}}} }{n^2(4l^2-n^2)} 
\end{eqnarray}
where $\tau_{\rm D}=L^2/D\pi^2$ is the dwell time, and $r=W/L$ is the aspect ratio.
For large aspect ratios $r$, the summation over $k$ can be replaced by an integration,
\begin{equation}\eqalign{
 \sum_{k=0}^{\infty} \frac{1}{l^2+k^2/r^2}  e^{-(k^2/r^2 )\frac{\tau_{\rm tr}}{\tau_{\rm D}}} &\approx r \int_0^{\infty} dx \frac{1}{l^2+x^2}e^{-x^2 \frac{\tau_{\rm tr}}{\tau_{\rm D}}}\nonumber\\
 &=r\frac{\pi}{2l}+O\left(\sqrt{\frac{\tau_{\rm tr}}{\tau_{\rm D}}}\right).
}\end{equation}
We are interested in leading terms in the small parameter ${\tau_{\rm tr}}/{\tau_{\rm D}}$ only, for which one finds 
\begin{equation}
  \delta G_{\rm WAL}=\frac{e^2 d_{\rm g}}{2\pi\hbar}\frac{W}{L}\frac{32}{\pi^3}\sum_{n=1,\, n\,\mathrm{odd}}^{\infty} \sum_{l=1}^{\infty}  
     \frac{l\,e^{-n^2 \frac{\tau_{\rm E}}{\tau_{\rm D}}} e^{-l^2 \frac{\tau_{\rm tr}}{\tau_{\rm D}}} }{n^2(4l^2-n^2)}.
\end{equation}
In the limit of small ${\tau_{\rm tr}}/{\tau_{\rm D}}$, the behavior of the summand for large $l$ is relevant. We then may simplify the summation over $n$ as follows: The main contributions arise for $n\approx 0$ and $n\approx 2l$, where the summand has poles. If $l$ is large, the poles are well separated, and the dominant contribution comes from the pole at $n\approx 0$,
\begin{equation}\eqalign{
 \sum_{n=1,\, n\,\mathrm{odd}}^{\infty}  
     \frac{1}{n^2(4l^2-n^2)}e^{-n^2 \frac{\tau_{\rm E}}{\tau_{\rm D}}}
    \approx \frac{1}{4l^2}\sum_{n=1,\, n\,\mathrm{odd}}^{\infty}  
     \frac{1}{n^2}e^{-n^2 \frac{\tau_{\rm E}}{\tau_{\rm D}}},
}\end{equation}
which results in the expression
\begin{equation}
  \delta G_{\rm WAL}=g\frac{e^2}{2\pi\hbar}\frac{W}{L}\frac{8}{\pi^3}\sum_{n=1,\, n\,\mathrm{odd}}^{\infty} \frac{1}{n^2} e^{-n^2 \frac{\tau_{\rm E}}{\tau_{\rm D}}}  \sum_{l}^{\infty}  
     \frac{1}{l} e^{-l^2 \frac{\tau_{\rm tr}}{\tau_{\rm D}}}. 
\end{equation}
For small $l$, this expression is not accurate, hence this summation has a lower cutoff, which is not relevant for small ${\tau_{\rm tr}}/{\tau_{\rm D}}$, however, where the $l$-summation results in $\ln \sqrt{{\tau_{\rm D}}/{\tau_{\rm tr}}}$. Hence, we find Eq.\ \eref{eq:WALresult} from the main text.

\section{Dephasing: Perturbation theory}

An alternative derivation of the dephasing correction to weak antilocalization can be obtained directly from perturbation theory in the interaction. Following \cite{aleiner1999}, to leading order in interaction, one finds two corrections to the conductance. The first one of these corresponds to the Altshuler-Aronov correction and was considered in Sec.\ \ref{sec:AA}. The second correction reads
\begin{eqnarray}
 \label{eq:dephasing}
 \fl \delta G_{\rm deph} 
  =
  -d_{\rm g}\frac{e^2 \hbar}{2\pi} \int dy \int dy' \int d\varepsilon \left(-\frac{\partial f(\varepsilon)}{\partial\varepsilon}\right)
\int \frac{d\omega}{2\pi} \left[\coth \left(\frac{\omega}{2 T}\right)-\tanh \left(\frac{\omega-\varepsilon}{2 T}\right)\right]
  \nonumber \\  \mbox{} 
\fl \qquad \qquad \times
\int d\vr_1 d\vr_2 \mathrm{Im} \left[U^{R}(\vr_1,\vr_2;\omega)\right]
  \nonumber\\  \mbox{}
  \fl \qquad \qquad \times
  \mathrm{tr}\left[\hat{v}_{x}{\cal G}^{\rm R}(\vr,\vr_1;\varepsilon) {\cal G}^{\rm R}(\vr_1,\vr_2;\varepsilon-\omega) {\cal G}^{\rm R}(\vr_2,\vr';\varepsilon)\hat{v}'_{x} {\cal G}^{\rm A}(\vr',\vr;\varepsilon)\right.\nonumber\\
 \fl  \qquad \qquad \qquad \mbox{} +\hat{v}_{x}{\cal G}^{\rm R}(\vr,\vr';\varepsilon) \hat{v}'_{x}{\cal G}^{\rm A}(\vr',\vr_2;\varepsilon) {\cal G}^{\rm A}(\vr_2,\vr_1;\varepsilon-\omega) {\cal G}^{\rm A}(\vr_1,\vr;\varepsilon)\\
 \fl  \qquad \qquad \left.\qquad \mbox{} +\hat{v}_{x}{\cal G}^{\rm R}(\vr,\vr_1;\varepsilon) {\cal G}^{\rm R}(\vr_1,\vr';\varepsilon-\omega)\hat{v}'_{x} {\cal G}^{\rm A}(\vr',\vr_2;\varepsilon-\omega) {\cal G}^{\rm A}(\vr_2,\vr;\varepsilon)  \right]_{x'=0, \, x=L}.\nonumber
\end{eqnarray}
The calculation proceeds by inserting the semiclassical expressions for the Green functions and identifying the relevant configurations of trajectories. Only configurations where ``advanced'' and ``retarded'' trajectories are paired up (where we also allow for small angle encounters or pairing of time-reversed trajectories) contribute systematically to the conductance. For the first term inside the trace in Eq.\ \eref{eq:dephasing}, this is only possible, if the three ``retarded'' trajectories join together to a single trajectory connecting the points $\vr'$ with $\vr$, that can be paired up with the advanced trajectory. In the semiclassical approximation, we then evaluate the integration over $\vr_1$ and $\vr_2$ within stationary phase approximation, where we keep only stationary configurations that join to a single trajectory. The result of such a calculation is
 \begin{eqnarray}
 \label{eq:convolution}
  \fl\int d\vr_1 d\vr_2  {\cal G}^{\rm R}(\vr,\vr_1;\varepsilon) {\cal G}^{\rm R}(\vr_1,\vr_2;\varepsilon-\omega) {\cal G}^{\rm R}(\vr_2,\vr';\varepsilon)
  \Im \left[U^{\rm R}(\vr_1,\vr_2;\omega)\right]
  ~~~~~ \nonumber \\
   =-\frac{1}{\hbar^2} \frac{2\pi}{(2\pi i\hbar)^{3/2}} \sum_{\alpha:\vr'\rightarrow\vr;\varepsilon} A_{\alpha} e^{i \mathcal{S}_{\alpha}/\hbar } |\chi(\vp_{\alpha})\rangle\langle\chi(\vp'_{\alpha})|e^{i\gamma_{\alpha}}
   \nonumber\\
  \quad \times  \int_{0}^{\tau_{\alpha}} dt \int_0^{t} dt' 
  \Im \left[U^{\rm R}(\vr_{\alpha}(t),\vr_{\alpha}(t');\omega)\right] e^{-i\omega(t-t')/\hbar}, ~~~~
\end{eqnarray}
We here restrict ourselves to explain, how the structure of this result can be understood, and refer to \cite{schneider2013} for the detailed calculation. The first step is to identify points $\vr_1$ and $\vr_2$, which make the total phase of the integrand stationary. Such configurations are obtained, whenever there exists a single classical trajectory $\alpha$ the connects the points $\vr'$ and $\vr$ via $\vr_2$ and $\vr_1$. Since the position of the intermediate points can be anywhere along the trajectory $\alpha$, the summation over stationary configurations of the intermediate points is expressed a summation over trajectories $\alpha$ as well as two time integrations along the trajectory $\alpha$. The Green function connecting the intermediate points is taken at a different energy, resulting in the additional factor $e^{-i\omega(t-t')/\hbar}$, as follows from Eq.\ \eref{eq:partialS}. Furthermore, the actions of the three subpaths sum up to the action $\mathcal{S}_{\alpha}$ of the joined 
path. Similarly, 
the individual 
Berry phases for the subpaths combine to the Berry phase of the joined path $\gamma_{\alpha}$, as the Berry phase is expressed as integral along the trajectory. Since the momenta are smoothly connected at the intermediate points, the intermediate spinors match together, and only the spinors at the endpoints remain in the final expression. The second step in the evaluation of the integral is to integrate over quadratic fluctuations around the stationary configurations. This, in turns, renders the proper stability amplitude $A_{\alpha}$ and the prefactor, see \cite{schneider2013}.

Similar considerations apply to the second and third term of the trace of  Eq.\ \eref{eq:dephasing}. Therefore, we can write Eq.\ \eref{eq:dephasing} as a sum over one ``retarded'' trajectory $\alpha$ and one ``advanced'' trajectory $\beta$, connecting source and drain. Since only paired trajectories are of relevance, we may simplify $A_{\alpha}=A_{\beta}$ and $\tau_{\alpha}=\tau_{\beta}$. Since the only dependence on the propagation energy $\varepsilon$ is in the factor between square brackets on the first line of Eq.\ (\ref{eq:dephasing}), we may perform the integration over $\varepsilon$ and find
\begin{equation}
 \int d\varepsilon \left(-\frac{\partial f(\varepsilon)}{\partial\varepsilon}\right) \left[\coth \left(\frac{\omega}{2 T}\right)-\tanh \left(\frac{\omega-\varepsilon}{2 T}\right)\right]=\frac{\frac{\omega}{2T}}{\sinh^2\frac{\omega}{2T}}.
\end{equation}
Inserting the Fourier transformed interaction we then obtain
\begin{equation}\eqalign{
 \label{eq:dephapp}
  \fl \delta G_{\rm deph}=& \frac{e^2 d_{\rm g}}{(2\pi\hbar)^2} \int dy \int dy' \sum_{\alpha,\beta} A_{\alpha}^2 v'_x v_x e^{i(\mathcal{S}_{\alpha}-\mathcal{S}_{\beta})} e^{i(\gamma_{\alpha}-\gamma_{\beta})}  
\nonumber\\ \fl &\times 
\int \frac{d\omega}{2\pi} \frac{\frac{\omega}{2T}}{\sinh^2\frac{\omega}{2T}}\int \frac{d\vq}{(2\pi)^2} \mathrm{Im} U^{R}(\vq;\omega)
\frac{1}{2\hbar^2}\int_0^{\tau_{\alpha}} dt_1 dt_2 e^{-i\omega(t_1-t_2)/\hbar} 
 \nonumber\\ \fl &\times 
\left[e^{i\vq\left[\vr_{\alpha}(t_1)-\vr_{\alpha}(t_2)\right]}+e^{i\vq\left[\vr_{\beta}(t_1)-\vr_{\beta}(t_2)\right]}-e^{i\vq\left[\vr_{\alpha}(t_1)-\vr_{\beta}(t_2)\right]}-e^{i\vq\left[\vr_{\beta}(t_1)-\vr_{\alpha}(t_2)\right]}\right],
}\end{equation}
which is consistent with the expressions of Sec.\ \ref{sec:Dephasing}.

\section{Dephasing: Loop segment}
\label{sec:appdeph}

In this appendix we add some details to the calculation of the dephasing for the loop segment.
The imaginary part of the screened interaction in the diffusive limit, Eq.\ \eref{eq:interactionfull}, evaluates to
\begin{equation}\eqalign{
\label{eq:ImUdiff}
\Im U^{R}(\vq,\omega)=&\Im U^{R,s}(\vq,\omega)+(d_{\rm g}^2-1) \Im U^{R,t}(\vq,\omega),
}\end{equation}
with
\begin{equation}\eqalign{
\Im U^{R,s}(\vq,\omega)=&-\frac{1}{d_{\rm g}\hbar \nu}\frac{ \omega}{ D q^2},\nonumber\\
 \Im U^{R,t}(\vq,\omega)=&-\frac{1}{d_{\rm g}\hbar \nu}\frac{(F_0^{\sigma})^2 \omega Dq^2}{(1+F_0^{\sigma})^2 (D q^2)^2+(\omega/\hbar)^2}.
}\end{equation}
Accordingly, we split Eq.\ \eref{eq:deltaP} as
\begin{equation}
 \delta P(\vr',\vr';t)=\delta P^{s}(\vr',\vr';t)+(d_{\rm g}^2-1)\delta P^{t}(\vr',\vr';t)
\end{equation}
Using Eq.\ \eref{eq:calP}, we are lead to the temporal integral
\begin{equation}\eqalign{
{\cal I}(\omega,\vq,t)=&
\int_0^t dt_1 dt_2
            e^{-i\omega(t_1-t_2)/\hbar} \nonumber\\
&\times\left[e^{-D q^2 \tau_2 (1-\tau_2/t)}-e^{-D q^2 \bar{\tau}_2 (1-\bar{\tau}_2/t)}\right],
}\end{equation}
where $\tau_2=|t_1-t_2|$ and $\bar{\tau}_2=|t-t_1-t_2|$. We evaluate this integral in the long-time limit $t\gg \max({\hbar}/{\omega},{1}/{Dq^2})$ (which is sufficient for the present analysis),
\begin{equation}\eqalign{
 \label{eq:timeint}
{\cal I}(\omega,\vq,t)
\simeq 2t\frac{Dq^2}{(Dq^2)^2+(\omega/\hbar)^2},
}\end{equation}
{\em i.e.}, we find a linear-in-$t$ behavior. 

For the $\vq$-integration we consider the non-singlet part of the interaction first,
\begin{eqnarray}
  \int \frac{d\vq}{(2\pi)^2} \frac{Dq^2\, \Im U^{R,t}(\vq,\omega)}{(Dq^2)^2+(\omega/\hbar)^2}
 &=&- \frac{\pi\hbar(F_0^{\sigma})^2}{4 g_0 (1+F_0^{\sigma})(2+F_0^{\sigma})}.
  \nonumber \\
\end{eqnarray}
Here we introduced the dimensionless conductance $g_0=2\pi\hbar d_{\rm g}\nu D$.
Our expression now reads
\begin{equation}
 \delta P^t(\vr',\vr';t)=-\frac{\pi t}{2g_0\hbar} \frac{(F_0^{\sigma})^2}{(1+F_0^{\sigma})(2+F_0^{\sigma})} \int \frac{d\omega}{2\pi} {\cal F}\left(\frac{\omega}{2T}\right),
\end{equation}
where ${\cal F}(x)=x/\sinh^2 x$. For small $\omega/T$, we may expand the function ${\cal F}(x)\approx 1/x$, which gives a logarithmic divergent $\omega$-integral. This integral should be cut at high energies by temperature, and at small energies by $\hbar/t$, where Eq.\ \eref{eq:timeint} ceases to be valid. So we find 
\begin{equation}
 \delta P^t(\vr',\vr';t)=- \frac{(F_0^{\sigma})^2}{(1+F_0^{\sigma})(2+F_0^{\sigma})} \frac{T t}{g_0\hbar} \ln\frac{T t}{\hbar}.
\end{equation}
The result for the singlet channel can be obtained from the latter equation by formally sending $F_0^{\sigma}\rightarrow\infty$.

\section*{References}

\providecommand{\newblock}{}

\end{document}